\documentclass[printer]{aa}
\usepackage{epsfig}
\begin{document}
\newcommand{\bea}{\begin{eqnarray}}    
\newcommand{\eea}{\end{eqnarray}}      
\newcommand{\be}{\begin{equation}}
\newcommand{\ee}{\end{equation}}
\newcommand{\bef}{\begin{figue}}
\newcommand{\eef}{\end{figure}}
\newcommand{\etal}{et al.}
\newcommand{\kms}{\,{\rm km}\;{\rm s}^{-1}}
\newcommand{\hubunits}{\,\kms\;{\rm Mpc}^{-1}}
\newcommand{\hmpc}{\,h^{-1}\;{\rm Mpc}}
\newcommand{\hkpc}{\,h^{-1}\;{\rm kpc}}
\newcommand{\msun}{M_\odot}
\newcommand{\K}{\,{\rm K}}
\newcommand{\cm}{{\rm cm}}
\newcommand{\cd}{{\langle n(r) \rangle_p}}
\newcommand{\Mpc}{{\rm Mpc}}
\newcommand{\kpc}{{\rm kpc}}
\newcommand{\xir}{{\xi(r)}}
\newcommand{\xrp}{{\xi(r_p,\pi)}}
\newcommand{\xsirpi}{{\xi(r_p,\pi)}}
\newcommand{\wrp}{{w_p(r_p)}}
\newcommand{\gr}{{g-r}}
\newcommand{\Navg}{N_{\rm avg}}
\newcommand{\Mmin}{M_{\rm min}}
\newcommand{\fiso}{f_{\rm iso}}
\newcommand{\Mr}{M_r}
\newcommand{\rp}{r_p}
\newcommand{\zmax}{z_{\rm max}}
\newcommand{\zmin}{z_{\rm min}}
\newcommand{\ve}[1]{\ensuremath{\mathbf{#1}}}
\newcommand{\D}[1][ ]{\ensuremath{\mathrm{d}^{#1} }}
\newcommand{\tdyn}{\ensuremath{\tau_\text{dyn}}}
\newcommand{\tve}[1]{\tilde{\boldsymbol{#1}}}

\def\eg{{e.g.}}
\def\ie{{i.e.}}
\def\spose#1{\hbox to 0pt{#1\hss}}
\def\ltapprox{\mathrel{\spose{\lower 3pt\hbox{$\mathchar"218$}}
\raise 2.0pt\hbox{$\mathchar"13C$}}}
\def\gtapprox{\mathrel{\spose{\lower 3pt\hbox{$\mathchar"218$}}
\raise 2.0pt\hbox{$\mathchar"13E$}}}
\def\inapprox{\mathrel{\spose{\lower 3pt\hbox{$\mathchar"218$}}
\raise 2.0pt\hbox{$\mathchar"232$}}}

\title{Extension and estimation of correlations in  Cold Dark Matter models}

\subtitle{}

\author{Francesco  Sylos Labini \inst{1,2} and  Nickolay L. Vasilyev \inst{3} } 

\titlerunning{Correlations in  CDM models}

\authorrunning{Sylos Labini \& Vasilyev } 

\institute{``Enrico Fermi Center'', Via Panisperna 89 A, 
Compendio del Viminale, 00184 Rome, Italy
\and``Istituto dei Sistemi Complessi'' CNR, 
Via dei Taurini 19, 00185 Rome, Italy
\and
Sobolev Astronomical Institute, St.Petersburg 
State University, Staryj Peterhoff, 198504,
St.Petersburg, Russia
}

\date{Received / Accepted}

\abstract{
We discuss the large scale properties of standard cold dark matter
cosmological models characterizing the main features of the
power-spectrum, of the two-point correlation function and of the mass
variance.  Both the real-space statistics have a very well defined
behavior on large enough scales, where their amplitudes become smaller
than unity.  The correlation function, in the range $0<\xi(r)<1$, is
characterized by a typical length-scale $r_c$, at which $\xi(r_c)=0$,
which is fixed by the physics of the early universe: beyond this scale
it becomes negative, going to zero with a tail proportional to
$-(r^{-4})$.  These anti-correlations represent thus an important
observational challenge to verify models in real space.  The same
length scale $r_c$ characterizes the behavior of the mass variance
which decays, for $r>r_c$, as $r^{-4}$, the fastest decay for any mass
distribution. The length-scale $r_c$ defines the maximum extension of
(positively correlated) structures in these models.  These are the
features expected for the dark matter field: galaxies, which represent
a biased field, however may have differences with respect to these
behaviors, which we analyze.  We then discuss the detectability of
these real space features by considering several estimators of the
two-point correlation function. By making tests on numerical
simulations we emphasize the important role of finite size effects
which should always be controlled for careful measurements.
}
\maketitle

\section{Introduction}

In contemporary cosmological models the structures observed today at
large scales in the distribution of galaxies in the universe are
explained by the dynamical evolution of purely self-gravitating matter
(dark matter) from an initial state with low amplitude density
fluctuations, the latter strongly constrained by satellite
observations of the fluctuations in the temperature of the cosmic
microwave background radiation.  The other main observational elements
for the understanding of the large scale structure of the universe is
represented by the studies of galaxy correlations. Any theoretical
model aiming to explain the formation of structures must be tested
against the data provided by galaxy surveys which give the important
bridge between the regimes characterized by large and small
fluctuations.

Models of the early universe (see e.g.  Padmanabhan, 1993 and
references therein) predict certain primordial fluctuations in the
matter density field, defining the correlations of the initial
conditions, i.e. at the time of decoupling between matter and
radiation.  In the regime where density fluctuations are small enough,
the correlation function of the present matter density field is simply
related to one describing the initial conditions. In fact, according
to the growth of gravitational instabilities in an expanding universe
in the linear regime perturbations are simply amplified (see e.g.,
Peebles, 1980 and references therein).  Thus today at some large
scales where the correlation function is still positive but with
$\xi(r)<1$ the imprint of primordial fluctuations should be
preserved. In the region of strong non-linear fluctuations an
analytical treatment to predict the behavior of the two-point
correlation function has not been developed yet and, in general, one
makes use of numerical simulations which provide a rich, but
phenomenological, description of structure in the non-linear
regime. It is in this regime, at small enough scales, where most
observations have been performed until now.

We focus here on the type of correlations predicted in the linear
regime by models of the early universe. While the characterization of
correlations is usually done in terms of the power-spectrum of the
density fluctuations a real space analysis turns out to be useful to
point out some relevant features from an observational point of view
(see, e.g., the discussion in Gabrielli et al., 2004).

Theoretical models of primordial matter density fields in the
expanding universe are characterized by a single well-defined length
scale, which is an imprint of the physics of the early universe at the
time of the decoupling between matter and radiation (see
e.g. Bond and Efstathiou 1984, and Padmanabhan 1993 for a general 
introduction to the problem). The redshift characterizing the
decoupling is directly related to the scale at which the change of
slope of the power-spectrum of matter density fluctuations $P(k)$
occurs, i.e. it defines the wavenumber $k_c$ at which there is the
turnover of the power-spectrum between a regime, at large enough $k$,
where it behaves as a negative power-law of the wave number $P(k) \sim
k^{m}$ with $-1<m\le-3$, and a regime at small $k$ where $P(k)\sim k$
as predicted by inflationary theories. Given the generality of this
prediction, it is clearly extremely important to look for this scale
in the data.

The exact location of this scale is related to several parameters,
including the cosmological ones which describe the geometry of the
universe at large scales (see e.g. Padmanabhan 1993, Tegmark et
al. 2004 and Spergel et al. 2007 for a recent determination). We
discuss in what follows that the scale $r_c$ corresponding to the
wave-number $k_c$, in a particular variant of Cold Dark Matter (CDM)
models --- the so-called $\Lambda$CDM vanilla model --- is predicted
to be $r_c
\approx 124$ Mpc/h\footnote{For
seek of clarity we have chosen the scale of distances normalized to
the adimensional Hubble parameter $h$, which is defined from the
Hubble's constant $H_0 = 100 h$ km/sec/Mpc.}. At this scale the real
space correlation function crosses zero, becoming negative at larger
scales. In particular the correlation function presents a positive
power-law behavior at scales $r \ll r_c$ and a negative power-law
behavior at scales $r \gg r_c$. Positive and negative correlations are
exactly balanced in way such that the integral over the whole space of
the correlation function is equal to zero. This is a global condition
on the system fluctuations which corresponds to the fact that the
distribution is super-homogeneous (or hyper-uniform),
i.e. characterized by a sort of stochastic order and by fluctuations
which are depressed with respect to, for example, a purely
uncorrelated distribution of matter (Gabrielli, Joyce and Sylos
Labini, 2002 --- see discussion below).

Note that the scale $r_c$ marks the maximum extension of
positively correlated structures: beyond $r_c$ the distribution must
be anti-correlated since the beginning, as the evolution time was not
sufficient for the positive correlations to be developed. Thus this
scale can be regarded as an upper limit to the maximum size of
structures (with large of weak correlations) in the present
universe. The possible discoveries of structures of larger size is
still a challenging task for observational cosmology.

A relevant problem for the measurements of small amplitude values of
the correlation function, i.e. when $\xi(r) <1$, is represented by the
characterization and the understanding of both the systematic biases
which may affect the estimators of $\xi(r)$ and the stochastic noise
which perturbs any real determination. A study of this problems can be
found, for example, in Kerscher (1999) and Kerscher et
al. (2000) where it is shown that in general the biases in several
estimators of the two-point correlation function are not
negligible. In particular when there are structures of large spatial
extension inside a given sample there can be non negligible biases
affecting the determination of two-point properties.  We focus here on
the systematic bias related to the effect of the so-called integral
constraint, which distorts any estimator of the correlation function
at large scales in any given sample. The integral-constraint
represents an overall condition on any estimator of the correlation
function which is due to the fact that the average density, estimated
in any given sample, is in general different from its ensemble average
value. 

Here we treat explicitly the case for the simplest estimator of the
two-point correlation function, {the so-called full-shell or minus
estimator and and we illustrate the situation for the other
estimatorsby studying artificial distributions. In particular we
devote most attention to the estimator introduced by Davis and Peebles
(1983), which is still very used in the literature, and to the
estimator introduced by Landy and Szalay (1993), which is the most
popular one. Kerscher et al. (2000) considered also other estimators,
like the Hewett estimator (Hewett, 1982) and the Hamilton estimator
(Hamilton, 1993) and have shown that the results obtained with he
Landy and Szalay estimator are almost indistinguishable from the
Hamilton estimator.

In this way we will be able to identify the problems related to the
identification of correlations above the mentioned scale $r_c$:
we will then propose several tests to be applied to the galaxy data,
in order to define the strategy to study the correlation function at
small amplitudes and larger distances in order to eventually
detect the length scale $r_c$.

Up to now studies of the correlation function $\xi(r)$ in galaxy
samples have been limited to small scales, i.e. $0.1 < r \ltapprox 30$
Mpc/h (i.e. Totsuji \& Kihara, 1969, Davis and Peebles, 1983,
Davis et al., 1988, Benoist et al., 1996 Park et al., 1994, Scranton
et al., 2002}, Zehavi et al., 2002, Zehavi et al., 2004,  Ross et
al., 2007) and only recently the volume covered by galaxy redshift
samples is approaching a size which is large enough to make a robust
estimation of the correlation function at scales of order 100 Mpc/h.
When the Sloan Digital Sky Survey (SDSS) (York et al., 2000) will be
completed by filling up the gap between the two main angular regions
of observations, which are nowadays disjointed, the volume of the
survey and the statistics of the number of objects in the samples
would be large enough to test space correlations on scales of order
$r_c$ or more. An exception to this situation is represented by the
paper by Eisenstein, et al., (2005), who, by studying a sample of
Luminous Red Galaxies (LRG) of the SDSS, have estimated the
correlation function on scales of order 100 Mpc/h. These authors have
however focused their attention to another real space feature of
theoretical models: the so-called ``bump'' of the correlation function
which corresponds in real space to the so-called Doppler peaks in the
matter power-spectrum generated by the baryonic acoustic oscillations
in the early universe. As we discuss below this bump, corresponding to
a singular point of the correlation function (Gabrielli et al., 2004),
is localized at scales of order of 100 Mpc/h and characterized by a
small amplitude. This is a second important real-space scale of the
theoretical correlation function which is localized at a scale
slightly smaller than $r_c$. The detection of the baryonic bump is
thus related to the detection of the scale $r_c$ as any finite-size
effect perturbing the determination of the scale $r_c$ will,
inevitably, also affect the determination of the baryonic bump.  In
fact the baryonic bump can be seen as a small modification to the
overall shape of the correlation function at scales of order $r_c$, to
which we focus our attention here.

Note that, because of the very large scales, the acoustic signature
and the zero point scale remain in the linear regime even today and
they are weakly affected by non-linear effects (see Eisenstein et al.,
2006). Thus real space and redshift space properties, at such large
scales, should not differ substantially.

In Section 2 we introduce the basic definitions of the statistical
quantities usually employed to characterize two-point properties in 
real and Fourier space. In Section 3 we discuss a simple functional
behavior of the power-spectrum of matter density fluctuations which
captures the main elements of a more realistic CDM power-spectrum. We
discuss the real-space properties as represented by the two-point
correlation function and we consider the problem of selection or
biasing in the simplest theoretical scheme of biasing a correlated
Gaussian field. In Section 4 we treat explicitly the case of a
$\Lambda$CDM matter density field characterizing in detail real space
properties.  The main estimators of the two-point correlation function
are discussed in Section 5 and in Section 6 we test these estimators
in artificial distributions.  Finally in Section 7 we draw our main
conclusions discussing the problems related to the estimations of
two-point correlations in real galaxy samples.

\section{Basic definitions} 

The microscopic number density function for any particle\footnote{We
make explicit the fact that we consider particle
distributions. However most of the definitions given hereafter can be
easily extended to the of a continuous matter density field. We refer
to Gabrielli et al.,  (2004) for more
details.}  distribution is given by
\begin{equation}
n(\ve x) = \sum_{i=1}^N \delta_D\left(\ve x- \ve x_i \right) \;,
\end{equation}
where $\ve x_i$ is the position of the $i$-th particle, 
$\delta_D$ is the Dirac delta function and the sum 
is over the $N$ particles of the system.

For a system in which the mean density $n_0$ is well defined and
positive, it is convenient to define the density contrast:
\begin{equation}
\delta(\ve x) = \frac{n(\ve x) - n_0}{n_0} \;.
\end{equation}
In order to characterize the two-point correlation properties of the
density fluctuations, one can then use the reduced two-point 
correlation function (hereafter simply  two-point 
correlation function):
\begin{equation} 
\tilde \xi(\ve r)= \langle \delta(\ve x+\ve r)\delta(\ve x)\rangle
 \;,
\end{equation}
where $\langle ... \rangle$ is the ensemble average, i.e., an average
over all possible realizations of the system. In a distribution of
discrete particles $\tilde{\xi}(\ve r)$ always has a Dirac delta
function singularity at $\ve r= \ve 0$, which it is convenient to
separate by defining $\xi(\ve r)$ for $\ve r \neq \ve 0$ (the
``off-diagonal'' part --- see e.g. Peebles 1980) 
\begin{equation}
\tilde{\xi}(\ve r) = \frac{1}{n_0} \delta_D(\ve r) + \xi(\ve r) \ . 
\label{eq:xigeneral}
\end{equation}

The normalized variance of particle number (or mass) is an integrated
quantity defined as :
\begin{equation}
\label{eq:variance} 
\sigma^2(r) = \frac{\langle N^2(r)\rangle - \langle N(r)\rangle^2
}{\langle N(r) \rangle^2} 
\end{equation}
where $N(r)$ is the number of particles inside, for example, a sphere
of radius $r$.  Then $\sigma^2(r)$ can be used, in a manner similar to
$\tilde \xi(\ve r)$, to distinguish a regime of large fluctuations
($\sigma^2 >1$) from a regime of small fluctuations where $\sigma^2
<1$. It is simple to find the explicit expression for the normalized
variance of particle number in terms of a double integral of $\tilde
\xi(\ve r)$ (see,  e.g., Peebles, 1980)
\be
\label{toy6}
\sigma^2(V)  = \frac{1}{V^2} \int_V\int_V \tilde \xi(|\vec{r_1}-\vec{r_2}|) d
^3r_1 d^3r_2 \;.
\ee

If we consider distributions which are periodic in a cube of side $L$,
we can write the density contrast as a Fourier series:
\begin{equation}
\delta(\ve x) = \frac{1}{L^3}\sum_{\ve k} \exp(i\ve k\cdot \ve x)
\,\tilde\delta(\ve k)
\label{eq:fourierdelta}
\end{equation}
with $\ve k \in \left\{ (2\pi/L) \ve n \, |\, \ve n \in
Z^3\right\}$.  The coefficients 
$\tilde\delta(\ve k)$ are given by
\begin{equation}
\tilde\delta(\ve k) =\int_{L^3} \delta(\ve x)
\exp(-i\ve k\cdot \ve x) \ \D[3] \ve x \ . 
\end{equation}
The power-spectrum  of a particle distribution is then defined (see
e.g., Peebles, 1980) as
\begin{equation}
P(\ve k) = \frac{1}{L^3}\langle | \tilde\delta(\ve k) |^2 \rangle \;.
\label{eq:pktheo}
\end{equation}
In point distributions which are statistically homogeneous, the
power-spectrum and the non-diagonal part of the  two-point
correlation function $\xi (\ve r)$ are a Fourier conjugate pair:
\be
\label{toy2}
\xi(\ve r) = \frac{1}{(2\pi)^3}\int d^3k P(\ve k) \exp(-i\ve k \ve r)
\ee
and 
\be
\label{pkinv}
P(\ve k)= \int d^3r  \xi(\ve r ) \exp(i\ve k \ve r ) \;.
\ee
Since for both $\xi(\ve {r})$ and $P(\ve {k}) $ we consider only the
dependence on the modulus of their arguments, we will denote them from
now on as $\xi(r)$ and $P(k) $ to mean that they are obtained by
performing an average over the directions of $\ve {r}$ and $\ve {k}$
respectively.

\section{A toy model and the problem of sampling}

In order to illustrate some key features of standard cosmological
models, let us consider a simple matter density field power-spectrum
of the type:
\be
\label{toy1}
P(k)=A k \exp(-k/k_c) \;.
\ee
This is characterized by an amplitude $A$ which fixes the small $k$ behavior
and by the turnover scale $k_c$ (see Fig.\ref{figps1}). 
\begin{figure}
\begin{center}
\includegraphics*[angle=0, width=0.5\textwidth]{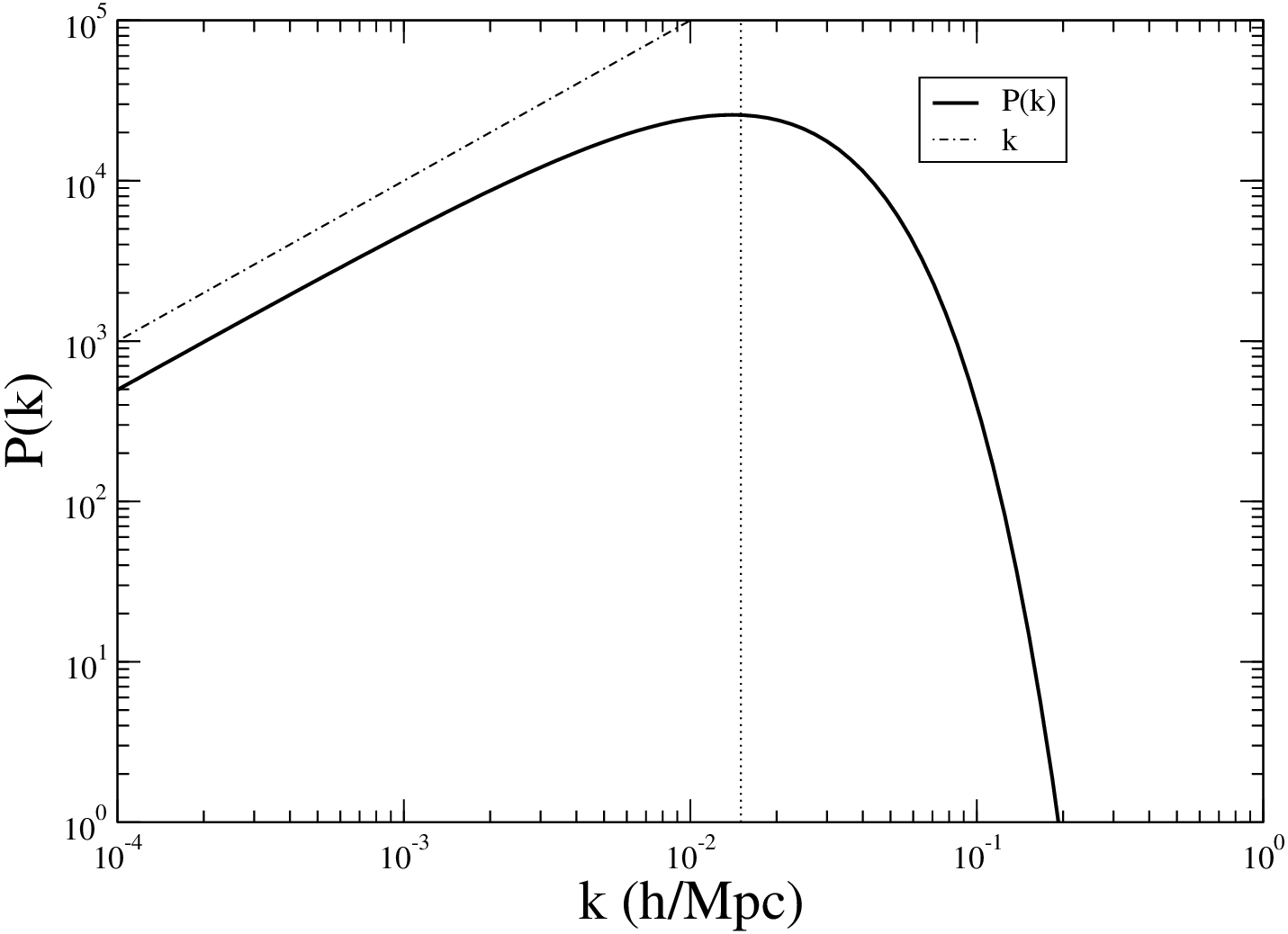}
\end{center}
\caption{Power-spectrum given by Eq.\ref{toy1}.  
The linear behavior at small $k$ is reported as a reference. The
amplitude at small $k$ and the scale $k_c=0.014$ h/Mpc are chosen to
be the same of the $\Lambda$CDM models discussed in what follows. The
vertical lines indicates the wave-length $k_c$.}
\label{figps1}
\end{figure}

As already mentioned, the two-point correlation function is simply the
Fourier transformation (FT) of the power-spectrum: for Eq.\ref{toy1}
by using Eq.\ref{toy6} we find
\be
\label{toy3}
\xi(r)= \frac{A}{\pi^2} \frac{\left( \frac{3}{k_c^2} - r^2\right)}
{\left( \frac{1}{k_c^2}+r^2\right)^3}  \;.
\ee
This correlation function presents the zero point at the intrinsic 
characteristic scale 
\be
\label{toy4} 
r_{c} = \sqrt{3}/k_c \;.
\ee
At small scales $r\ll r_c$ Eq.\ref{toy3} gives $\xi(r) \approx$
const. $> 0 $; while at large scales $r\gg r_c$ the amplitude of $\xi(r)$ becomes
negative, going to zero for $r\rightarrow \infty$ with a power-law tail
of the type $\xi(r) \approx -r^{-4}$ (see Fig.\ref{figxi1}).
\begin{figure}
\begin{center}
\includegraphics*[angle=0, width=0.5\textwidth]{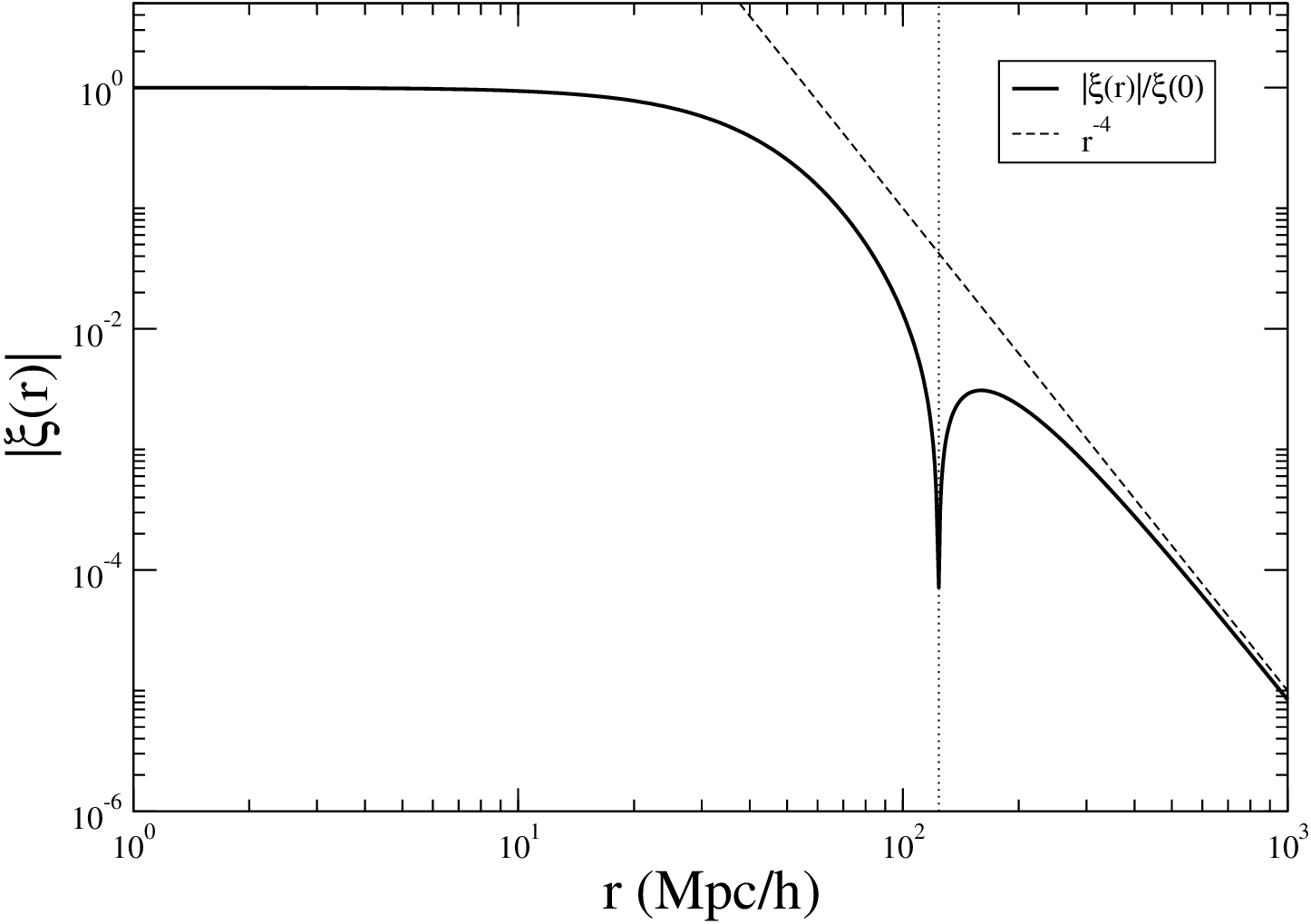}
\end{center}
\caption{Absolute value of the two-point correlation 
function given by Eq.\ref{toy3} divided by $\xi(0)$.  The (negative)
power-law $r^{-4}$ is shown as a reference. The vertical lines indicates the scale $r_c$.}
\label{figxi1}
\end{figure}

The region of positive correlation is thus followed by an (infinite)
region where there are anti-correlations. Positive and negative
correlations are exactly balanced so that 
\be
\label{toy4b}
\int_0^{\infty} \xi(r) r^2 dr = 0 \;.
\ee
This is equivalent to the condition that $P(k) \rightarrow 0$ for
$k\rightarrow 0$.  As discussed in Gabrielli, Joyce, Sylos Labini 
(2002) (see also Gabrielli et al., 2004) this corresponds to the
fact that the distribution is globally super-homogeneous, i.e. more
ordered than an uncorrelated distribution (i.e. a Poisson). This
subtle property can be clarified by computing the mass variance.

To evaluate the mass variance (Eq.\ref{toy6}) one may choose as the
volume of integration $V$ a sphere in real space of radius $R$.  In
this case, going into Fourier space, Eq.\ref{toy6} becomes (see
e.g., Peebles, 1980)
\be
\label{toy7}
\sigma^2(R)  = \frac{9}{2\pi^2} 
\int_0^{\infty} dk  k^2 P(k) 
\frac{\left( \sin(kR) + (kR)\cos(kR)\right)^2}{(kR)^6} \;. 
\ee
By considering the power-spectrum given by Eq.\ref{toy1} one finds
that $\sigma^2(R) \approx$ const. for $R <r_c$ and $\sigma^2(R) \sim
R^{-4}$ for $R > r_c$ (see Fig.\ref{figvar1}). This fast decay of the
mass variance is the distinctive feature of super-homogeneous mass
distributions and it is strictly related to the condition $P(0)=0$.
This is the fastest decay possible for {\it any} isotropic
translationally invariant distribution of points (see discussion in
Gabrielli, Joyce and Sylos Labini, 2002).

For a Poisson distribution one finds that the mass variance decays
slower than for a super-homogeneous distribution, i.e. $\sigma^2(R)
\sim R^{-3}$, and that the power-spectrum obeys to
\be
\lim_{k\rightarrow 0} P(k)= \mbox{const.} >0 \;. 
\ee
A similar situation occurs in the case the distribution has positive
correlations at small scales and no correlations at large scales --- a
substantially Poisson distribution.  On the other hand, in the
presence of long-range positive correlations, as for example a power
law correlation function $\xi(r) \sim r^{-\gamma}$, with $0<\gamma<3$,
the mass variance decays slower than the Poisson case,
i.e. $\sigma^2(R)
\sim R^{(\gamma-3)}$, and the power-spectrum satisfies the condition 
\be
\lim_{k\rightarrow 0} P(k)= \infty  \;.
\ee
\begin{figure}
\begin{center}
\includegraphics*[angle=0, width=0.5\textwidth]{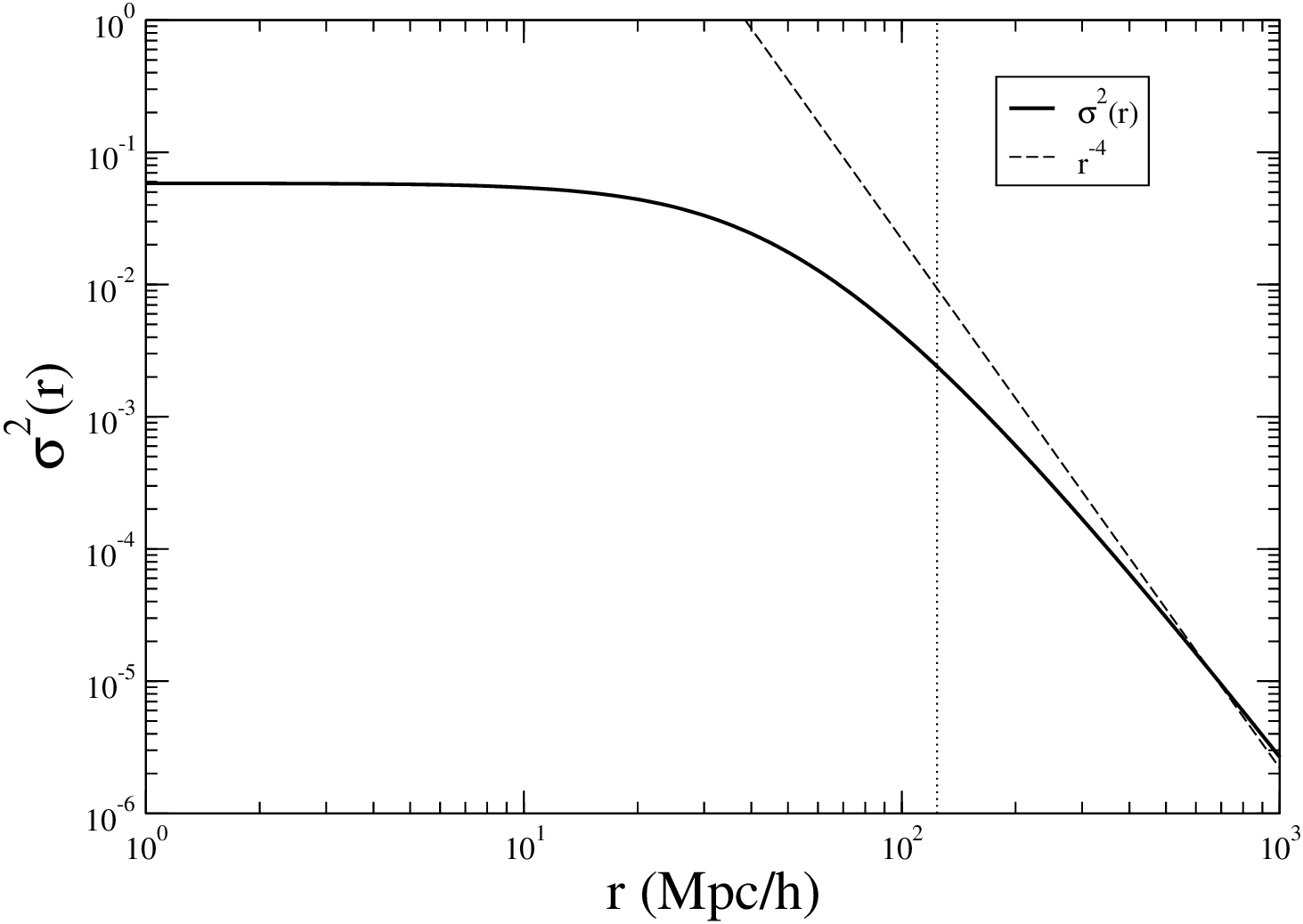}
\end{center}
\caption{Variance in real space spheres 
for Eq.\ref{toy1}. It is reported a line with slope $r^{-4}$ as
reference. The vertical lines indicates the scale $r_c$. }
\label{figvar1}
\end{figure}

\subsection{Sampling a density field} 
\label{sec_sampling}

What we have just described is a simple toy model power-spectrum which
captures some essential features of the theoretical correlation
properties of the matter density field. In the discussion of real
galaxy samples, one has to consider that luminous objects trace the
underlying dark matter density field and that they can be regarded as
a sampling of it: for example they can be supposed to lie in the
highest peaks of the fluctuations field, because only there
gravitational clustering has been efficient enough to form
self-gravitating objects.  The problem of sampling is thus a central
one in studies of cosmological density fields and, particularly, of
galaxy structures. More precisely by sampling we mean the operation
performed when one extracts, from a given distribution, a subsample of
it by using a selection criteria based on a certain parameter
characterizing the distribution. For example, one can make such type
of selection by extracting from the whole population of galaxies of
all luminosity, only those objects whose luminosity is brighter than a
given threshold; alternatively a similar selection can be done by
considering galaxy color. In the case the fluctuation field is a
stochastic variable of position (for example a Gaussian fluctuation
field) one may sample the distribution by selecting only fluctuations
larger than a given threshold in the density fluctuation field.

In general the problem consists in the understanding of the relations
between the statistical properties of the sampled, or biased,
distribution with those of the original one. A particular interest
lies in the relation between the two-point correlation function of the
sampled field with the original $\xi(r)$. This is so because, for
instance, in the studies of galaxy samples, one naturally has to
perform a sampling when measuring the two-point correlation function
of galaxies of a certain luminosity. In the comparison of observations
with theoretical models the sampling procedure is strictly related to
the physics of the system. In fact, in the analysis of cosmological
N-body simulations one also needs to extract subsamples of points
which, according to some models, would represent galaxies instead of
dark matter particles. In these contexts, the simplest theoretical
model describing biasing (introduced by Kaiser, 1984) was developed
for a continuous Gaussian field, and thus it does not represent an
useful analytical treatment of the problem of strong clustering, which
is instead the relevant one for galaxy structures.

However it is very difficult to treat the problem of sampling for a
generic case unless one may specify in detail the correlation
properties of the original distribution and the specific procedure
used to make the sampling.  This is a task which is out of current
knowledge even for the case of artificial distributions generated by
gravitational N-body simulations where one can make a phenomenological
approach.  For this reason, we limit the discussion to the threshold
sampling of the Gaussian random fields, because this allows us to
point out some key-features which characterize the case in which the
underlying density field has super-homogeneous type correlations and
the sampling is local (i.e.  related to local features of the
distribution).  This cannot be regarded as a realistic example for the
reasons discussed above, but one may identify several key problems
which should be addressed in detail by means of studies of artificial
distributions generated, for example, by N-body simulations for the
understanding of a more realistic case.

\subsection{Sampling a Gaussian random field} 
\label{sec_bias}

Let us now discuss the simplest biasing scheme of a continuous and
correlated Gaussian field (hereafter we follow Durrer et al., 2003).
Suppose to have a Gaussian random field with two-point correlation
$\xi(r)$ and such that the variance is $\langle \mu^2 \rangle =
\sigma^2$ (where $\mu$ is the mean density normalized 
fluctuation). One can identify fluctuations of the field such that
they are larger than $\nu$ times the variance. This selection defines
a biased field with the weight equal to zero if the fluctuations of
the original field are smaller than $\tilde \mu\equiv \nu\sigma$ and equal to
one if they are equal or larger than $\tilde \mu$. When one changes the
threshold $\nu$ one selects different regions of the underlying
Gaussian random field, corresponding to fluctuations of differing
amplitudes. The two-point correlation function of the selected
objects is then that of the peaks $ \xi_{\tilde \mu}(r)$.

We define the two-point correlation function of the normalized field
\be
\hat \xi(r) = \frac{\xi(r)}{\xi(0)} \,, 
\ee 
where $\xi(0)$ is the variance of the field so that $\hat \xi(r) \le 1
\;\; \forall r$. It is possible to compute the following first-order
approximation (Durrer et al., 2003)
\be
\xi_{\tilde \mu} (r)\approx 
\sqrt{\frac{1+\hat \xi(r)}{1-\hat \xi(r)}}\exp\left(\nu^2
\frac{\hat \xi(r)}{1+\hat\xi(r)}\right)-1
\label{approx-xinu} \;, 
\ee
which reduces to $\xi_{\tilde \mu} (r) \simeq \nu^2 \hat \xi(r)$ when $\nu^2
\hat |\xi(r)| \ll 1$.   Thus, if present in the underlying distribution,
 the characteristic length scale of the zero point $r_{c}$ is not
 changed under this selection procedure, i.e.
\be
\label{zpk}
\xi_{\tilde \mu} (r_{c})  = \hat \xi(r_{c})=0  \;\; \forall \tilde \mu \;.
\ee
On the other hand for $\xi_{\tilde \mu} (r) >1$ the amplification is non-linear
as a function of scale: this means that the functional behavior of
$\xi_{\tilde \mu}(r)$ is different from the one of $\hat \xi(r)$ in the regime
where $\xi_{\tilde \mu} (r) >1$.  Fig.\ref{xi.exp.bias} shows the situation when
one takes the correlation function of the toy model discussed in the
previous section (see Eq.\ref{toy3}) as the $\xi(r)$ of the underlying
Gaussian field.

\begin{figure}
\begin{center}
\includegraphics*[angle=0, width=0.5\textwidth]{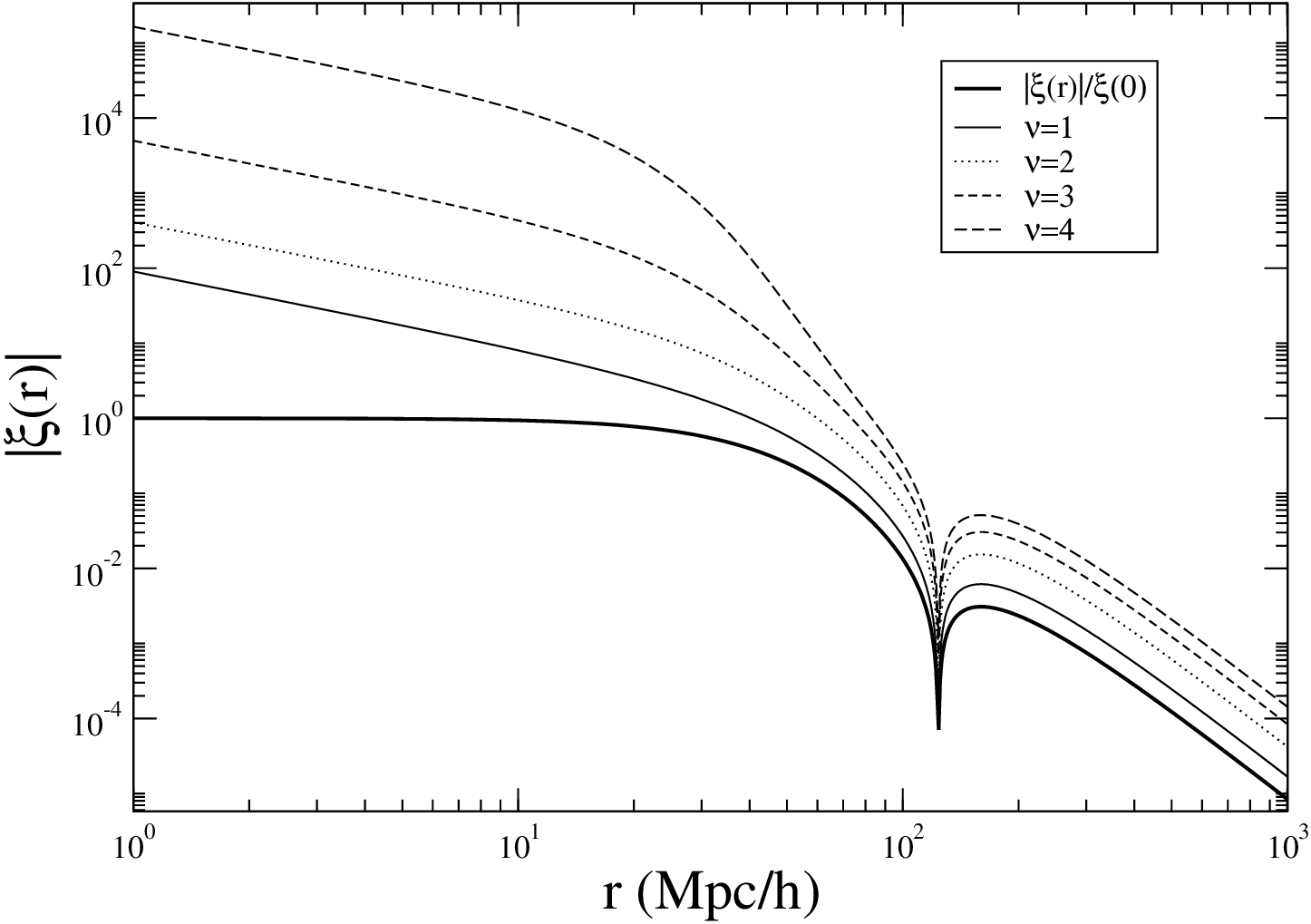}
\end{center}
\caption{Absolute value of the 
correlation function of the toy model described by Eq.\ref{toy1} 
(solid line) and of the ones corresponding to different values of the
threshold parameter $\nu$ calculated by applying
Eq.\ref{approx-xinu}. The amplification is non-linear at small scales,
where $\xi_{\tilde \mu}(r)  >1$, linear at large scales, and the
zero-crossing scale is invariant under biasing.}
\label{xi.exp.bias}
\end{figure}
Given the asymmetrical amplification at small and at large scales the
condition of super-homogeneity is broken, i.e.
\be
\int_0^{\infty} \xi_{\tilde \mu}(r) r^2 dr >0 \;, 
\ee
and thus the power-spectrum does not show anymore the tail $P(k)
\sim k$ (see Fig.\ref{pk.exp.bias}).
\begin{figure}
\begin{center}
\includegraphics*[angle=0, width=0.5\textwidth]{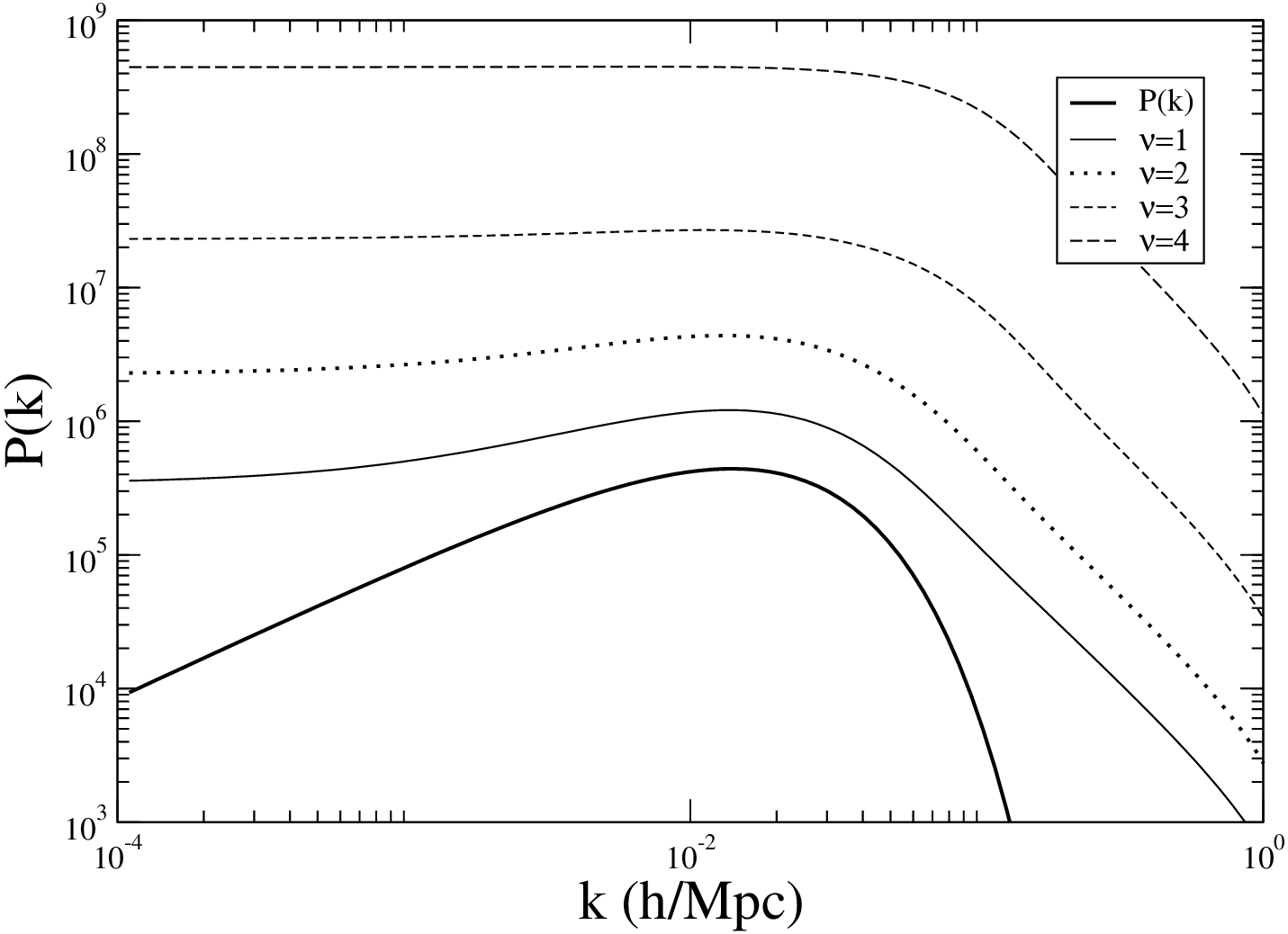}
\end{center}
\caption{ Power-spectrum of the toy model described by Eq.\ref{toy1} 
(solid line) and of the ones corresponding to different values of the
threshold parameter $\nu$ calculated by applying Eq.\ref{approx-xinu}
and then by making the Fourier Transformation. Because of the
asymmetrical amplification of the correlation function at small and at
large scale the condition of super-homogeneity is broken,  i.e. the
power-spectrum does not show anymore the tail $P(k) \sim k$. }
\label{pk.exp.bias}
\end{figure}
Correspondingly the mass variance shows the typical features of a
substantially Poisson system beyond the scale $r_c$, i.e. it decays as
$r^{-3}$.

Summarizing the behaviors for the toy model described by Eq.\ref{toy1} 
we obtain that:
\begin{itemize} 
\item  (i)
the correlation function of the biased field still presents some key
features of the original correlation function, namely the same
characteristic scale $r_c$ and the same negative tail $\xi(r) \sim
-r^{-4}$ at large scales. 
\item (ii) The power-spectrum is  distorted in a non-linear way at all scales by
biasing; in particular at large scales this is characterized by the
typical behavior of a Poisson distribution.  The same situation
occurs for the mass variance.
\end{itemize} 
 
We expect these to be general features of the biased fields when the
underlying density field has super-homogeneous type correlations
(Durrer et al. 2003, Gabrielli, et al., 2004). The cancellation of the
super-homogeneous features is due to the fact that the operation of
selection introduces a noise, due to the sampling itself, which
dominates the intrinsic fluctuations of the system.

\section{Real space correlations in CDM-type models}

In this section we consider the case of a distribution with
correlation properties of CDM type. In particular we study the case of
the so-called $\Lambda$CDM ``vanilla'' model. The functional behavior
and the parameters defining this model are discussed in Tegmark et
al. (2004) and Spergel et al., (2007).  Without entering into the
details of the model here we note that while the different
cosmological parameters may change the behavior of the power-spectrum
in a non-linear way, it is generally assumed that the bias factor $b$
(we now indicate by $b$ what in the previous section we have called
$\nu$ in order to make clear that the latter symbol refers only to the
case of a correlated Gaussian field) corresponds to an overall rescale
of its amplitude:
\be
\label{psbias} 
P(k)=b^2 P_{dm} (k)
\ee
where $P_{dm} (k)$ represents the power-spectrum of the underlying
dark matter field and $P(k)$ is the ``biased'' power-spectrum,
corresponding to the power-spectrum of a field selected by following a certain
prescription. As discussed above Eq.\ref{psbias} does not have any
theoretical justification in the framework of Gaussian fields neither
at small $k$ nor at large $k$. Rather in numerical simulations it has
been phenomenologically found that this is a good working hypothesis
in the regime of strong clustering (Springel et al., 2005).

In order to compute the real space properties it is useful to find an
analytical approximation to the theoretical power-spectrum which can
be found numerically (we use hereafter the data from Tegmark et al.,
2004). We have found that the following expression provides us with a
good fitting formula 
\be
\label{k1} 
P(k)=\frac{A k}{(1+B(k/k_1)^{\nu_1}+(k/k_2)^{\nu_2})} 
\ee
where $A =5 \cdot 10^6$, $B=10^3, k_1=0.35$ h/Mpc, $ \nu_1=2.3$,
$k_2=0.05$ h/Mpc, $\nu_2=3.5$. This power-spectrum is characterized by
a turnover scale $k_c\approx 0.014$ h/Mpc which separates the large
scales behavior $P(k) \sim k$ from the small scales one $P(k) \sim
k^{-2}$. 

\begin{figure}
\begin{center}
\includegraphics*[angle=0, width=0.5\textwidth]{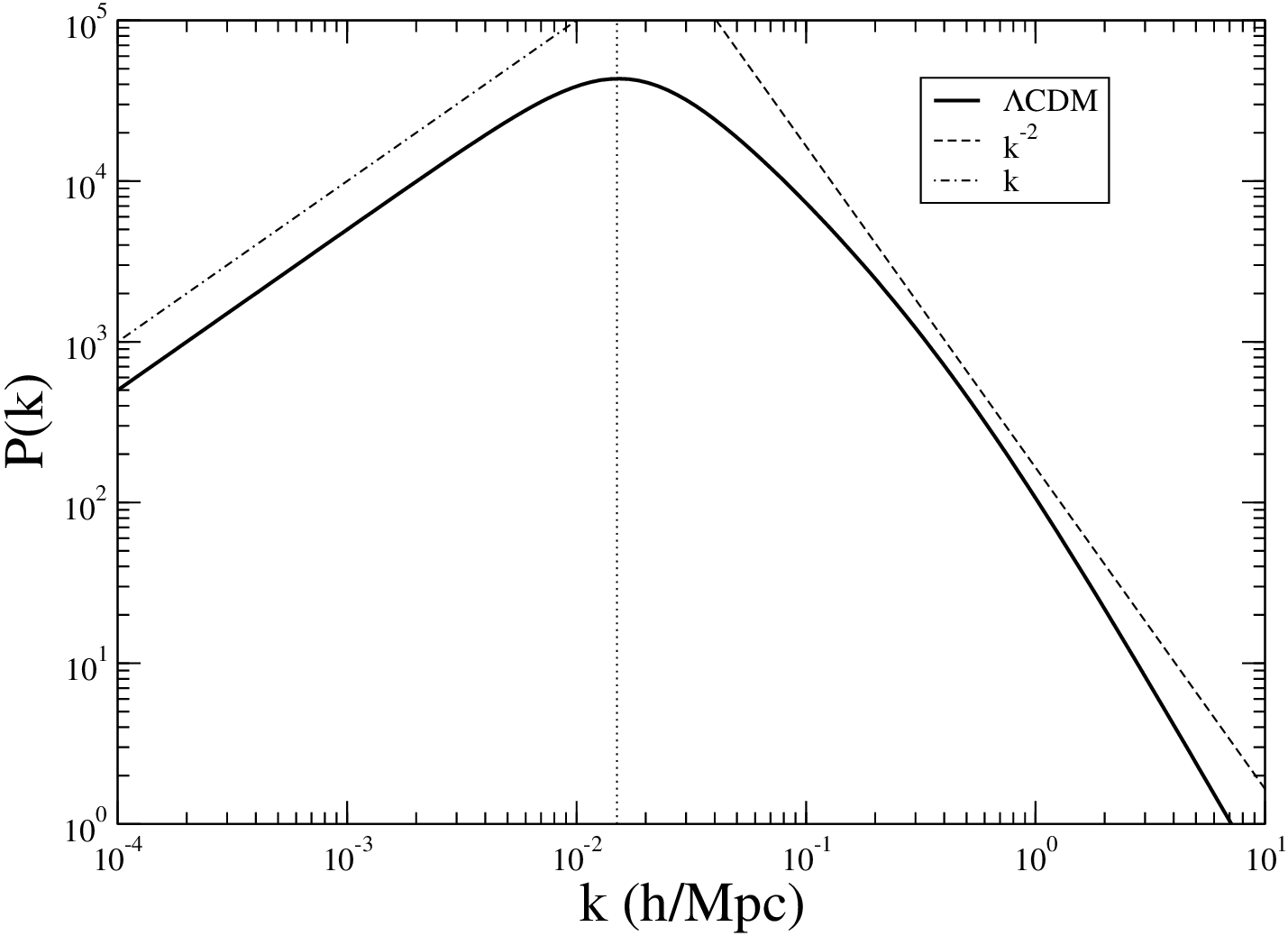}
\end{center}
\caption{Power-spectrum for the  $\Lambda$CDM  model (Eq.\ref{k1}). The two
power-laws $P(k) \sim k$ and $P(k) \sim k^{-2}$ are shown as a
reference.}
\label{figps}
\end{figure}

In this case it is not possible to calculate analytically the real
space correlation function, but it can be obtained from the numerical
computation of the Fourier transform of the power-spectrum by using
Eq.\ref{toy2}.  The result is shown in Fig.\ref{figxi}. As for the
case of the toy model discussed in the previous section, this
correlation function is characterized by a positive region at small
scales, where in this case it decays roughly as $r^{-1.5}$, and by a
large scale negative tail $\xi(r) \sim -r^{-4}$. The length scale
which separates these two regimes is the zero-point $r_c$ which
represents the unique characteristic length scale of this model: for
the parameters chosen in Eq.\ref{k1} we find $r_c=124$Mpc/h.

A reasonable fit to the correlation function obtained by making the FT
is (see Fig.\ref{figxi4})
\be
\label{xi2}
\xi(r)= \frac{A}{\pi^2} \frac{\left( \frac{3}{k_c^2} - r^2\right)}
{\left( \frac{1}{k_c^2}+r^2\right)^3} \cdot
\left( \frac{r^\beta+\frac{3^{(\beta-2)/2}}{k_c^\beta} } {r^\beta} \right)  \;
\ee
where $A=5 \cdot 10^6$ and $k_c=0.014$h/Mpc and $\beta=1.4$.

\begin{figure}
\begin{center}
\includegraphics*[angle=0, width=0.5\textwidth]{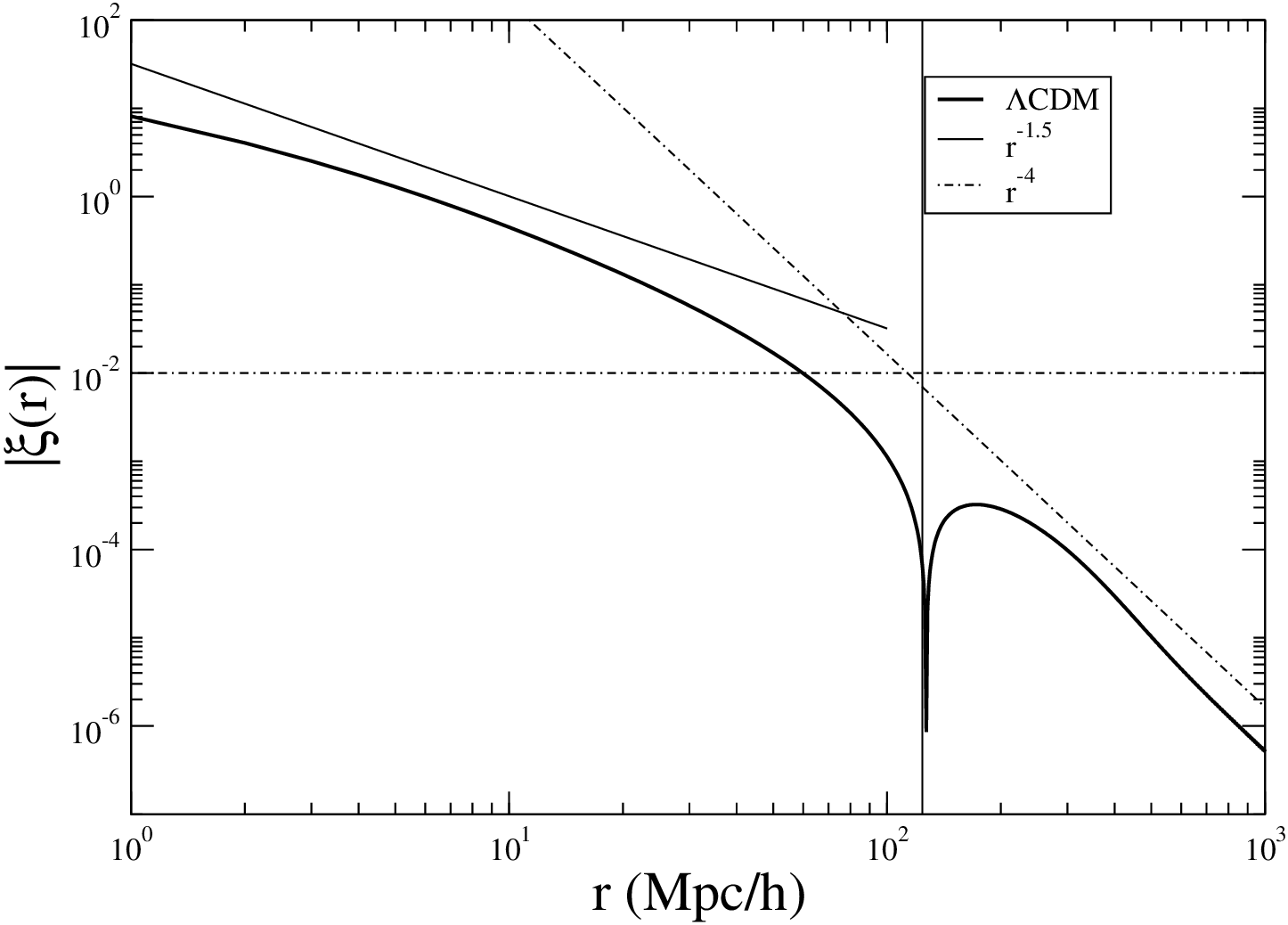}
\end{center}
\caption{Absolute value 
of the two-point correlation function for the $\Lambda$CDM  model
(Eq.\ref{k1}). The two power-laws $r^{-1.5}$ and $r^{-4}$ are shown as
a reference.}
\label{figxi}
\end{figure}
\begin{figure}
\begin{center}
\includegraphics*[angle=0, width=0.5\textwidth]{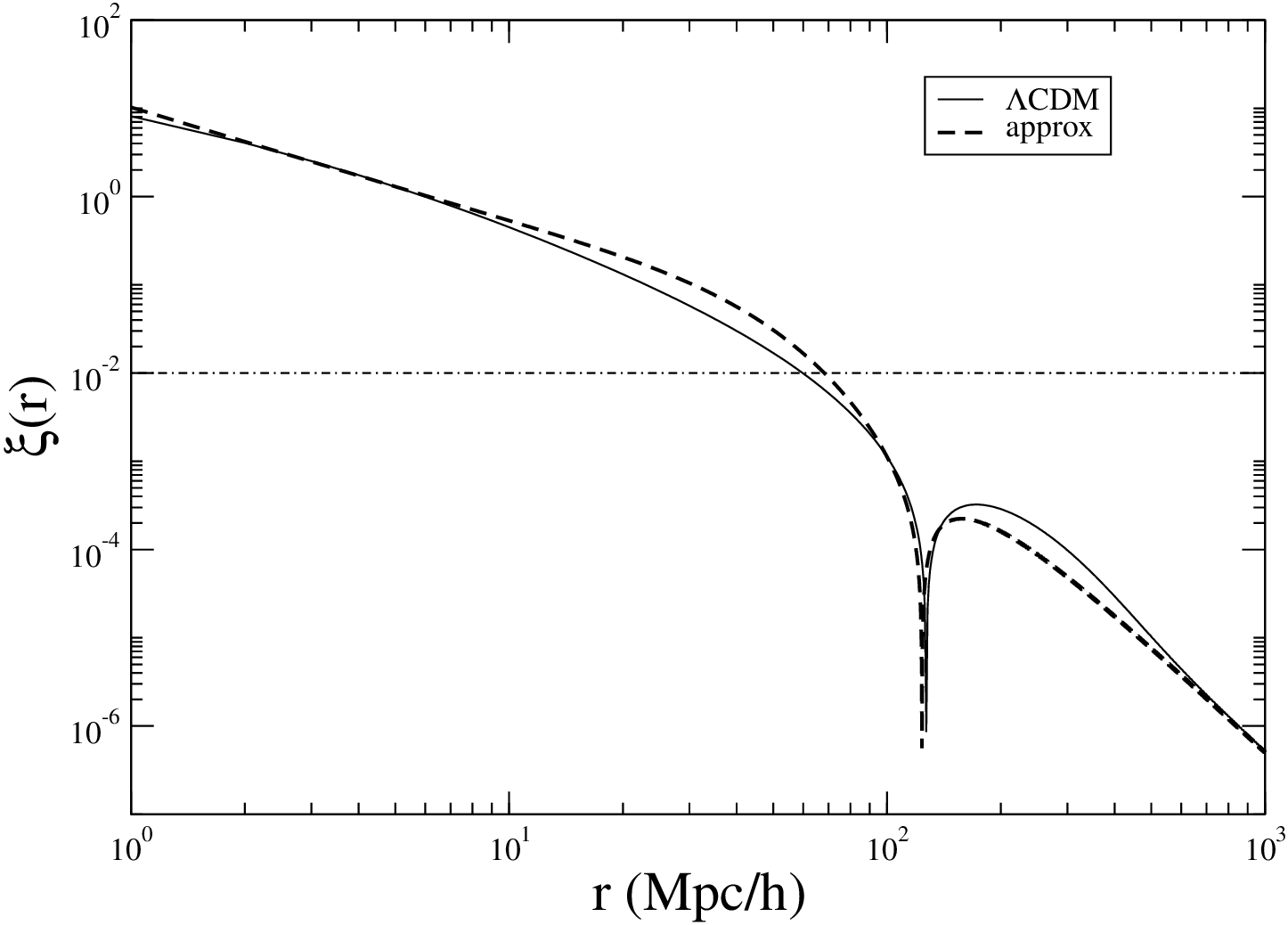}
\end{center}
\caption{Correlation function for the $\Lambda$CDM model (Eq.\ref{k1}) and 
the approximation given by Eq.\ref{xi2}}
\label{figxi4}
\end{figure}

It is interesting to note that if we compute the power-spectrum 
calculating the FT of the correlation function by using the analytical
approximation given by Eq.\ref{xi2}, although the fit is very good
over the all range of scales considered, we do not get the correct
behavior at small wave-modes, i.e. that $P(k) \sim k$ for $k < k_c$:
instead we get $P(k)\sim$ const. for $k<k_c$ (see Fig.\ref{figps2}).
\begin{figure}
\begin{center}
\includegraphics*[angle=0, width=0.5\textwidth]{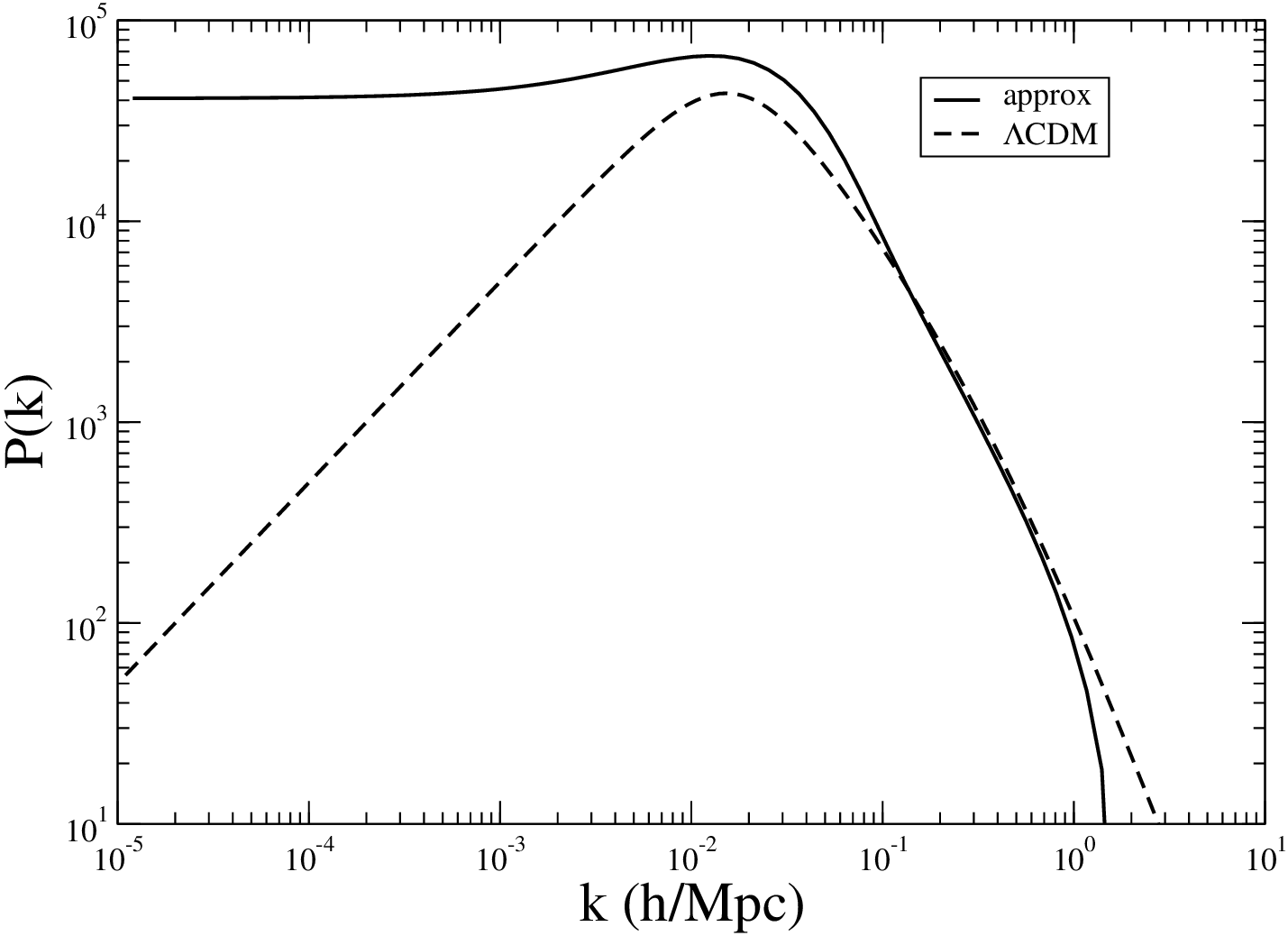}
\end{center}
\caption{Power-spectrum for the  $\Lambda$CDM model and the approximation 
obtained from Eq.\ref{pkinv}
by using Eq.\ref{xi2}.}
\label{figps2}
\end{figure}
This is because the small approximation introduced 
in Eq.\ref{xi2} is such that the integral
\be
\int_0^{\infty} \xi(r) r^2 dr > 0
\ee
and thus there is no the perfect cancellation between the positive and
negative parts, i.e. the typical feature of super-homogeneous
distributions, characterized by an extremely fine-tuning of the
correlations. This simple example shows how sensible is the condition
of super-homogeneity and gives a feeling of the kind of problems which
can arise in the framework of sampling. In general, the amplification
of the correlation function due to selection (or bias) is not linear
and gives rise to a behavior like one just described, i.e. to the
radical change of the super-homogeneous properties. That is, the
distribution becomes substantially Poisson on scales larger than $r_c$
because of the noise introduced by sampling, although the negative
$-r^{-4}$ tail in the correlation function is still present.

\subsection{Main features of the  real space two-point correlation function}

As discussed above, the regime of large fluctuations \mbox{$\xi(r)>1$}  is not
predictable by a theoretical approach, and thus both the amplitude and
the shape of the correlation function have to be constrained by
observations.  Any specific model of matter density field 
however predicts the behavior of the correlation function in the
regime $|\xi(r)|<1$. We discuss, as an interesting example, the case
of the  $\Lambda$CDM model mentioned above.

In general, it is possible to characterize the approach of the
correlation function to the zero point, in a range of scales such that
$0<\xi(r) $. For the case of the  $\Lambda$CDM model we get that in this
range of scales a good and useful approximation is given by
\be
\label{xiapprox}
\xi(r)\approx A \left(\frac{\lambda}{r}\right)^\gamma \exp(-r/\lambda) 
\ee
where $A=3 \cdot 10^{-1}$,  $\lambda=25$Mpc/h and $\gamma=1$, while $A=3
(0.03)$ when the amplitude of Eq.\ref{k1} is multiplied by a factor 10
(1/10).  The result is shown in Fig.\ref{figxi3}. The exponential
cut-off, independent on bias, is related to the fact that
$\xi(r)$ crosses zero at $r_c=124$ Mpc/h.
\begin{figure}
\begin{center}
\includegraphics*[angle=0, width=0.5\textwidth]{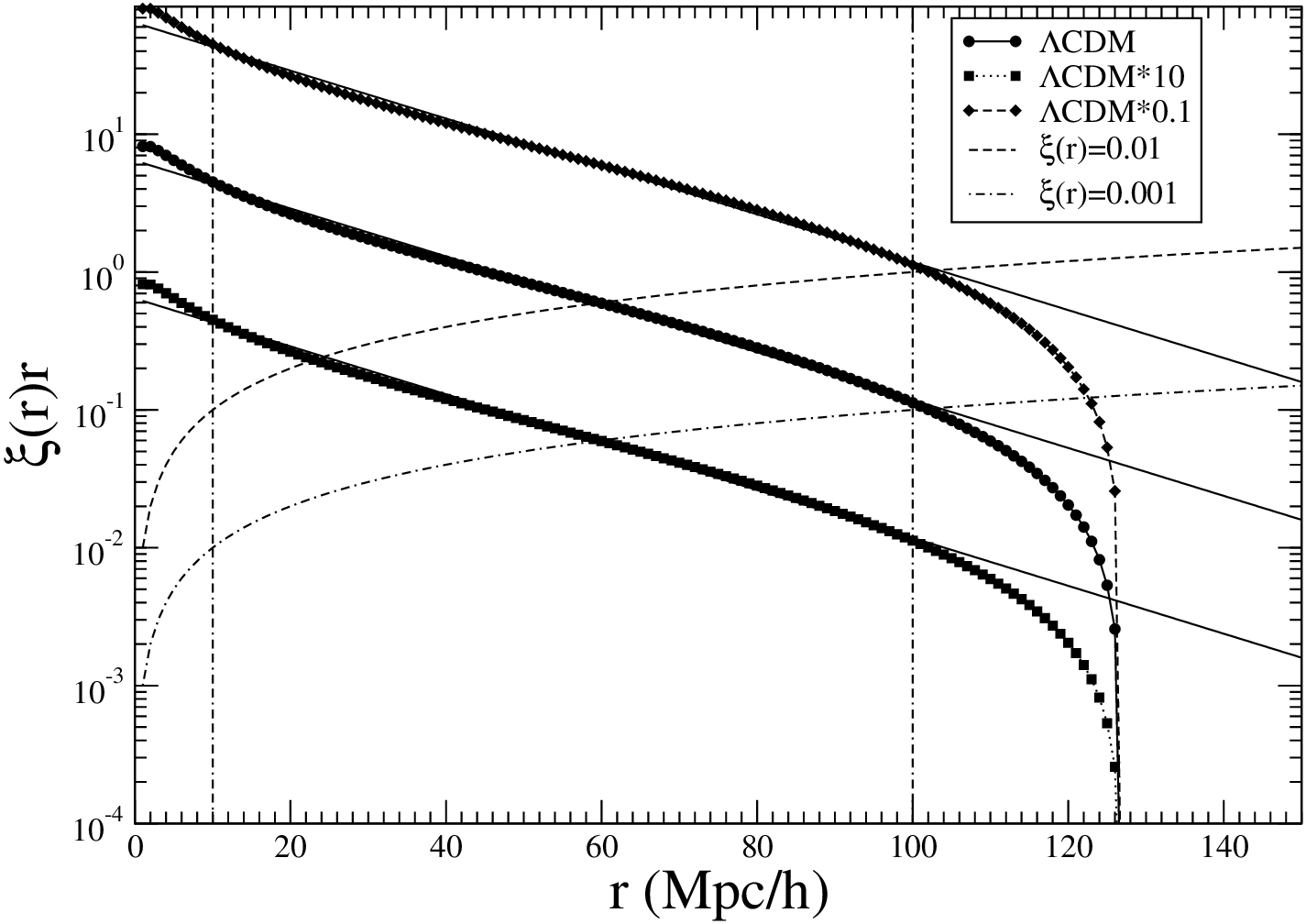}
\end{center}
\caption{Plot of the function $\xi(r) \times r$ 
for the  $\Lambda$CDM model (Eq.\ref{k1}) and, for comparison for the case
in which the amplitude of Eq.\ref{k1} as been multiplied by a factor
10 and a factor 1/10. The dashed lines correspond to the thresholds 
such that $\xi(r)=0.01,0.001$.}
\label{figxi3}
\end{figure}
Thus while the direct identification of the zero-point scale is
clearly very difficult in a finite sample (see discussion below), for
the effect of stochastic and systematic noise in the estimators, the
approach to the zero point, in this model, is very well defined. In
particular, the correlation function presents an exponential decay in
the range of scales [10,100] Mpc/h. Depending on the value of the
amplitude of $\xi(r)$, this range of scales is extended enough in the
region where $\xi(r) > a$ with $a>10^{-2}$, thus a region where maybe
observations will be provide with statistically robust samples, for a
bias factor of order one for the parameters considered here.

\subsection{The Baryonic Bump} 

As mentioned in the introduction, according to the physics of the
early universe sound waves propagating in the first $\sim$ 400,000
years after the Big Bang produce an additional characteristic length
scale in the matter and radiation density fields. With galaxy surveys
it would be possible to detect this acoustic feature as a bump in the
correlation function at $\sim$ 100 Mpc/h. The amplitude of this bump
is controlled by the baryon density, the matter density and the Hubble
constant (see Eisenstein et al., 2005 for a detailed discussion). It
is interesting to note that this bump corresponds to a non-analytical
point of the correlation function which gives rise to a co-sinusoidal
modulation for the power-spectrum (see Gabrielli et al., 2004).

In Fig.\ref{figbb} we show  a typical example. (In this the case
the matter density is $\Omega_m=0.12 h^{-2}$ and the baryon density is
$\Omega_b=0.024 h^{-2}$.)
\begin{figure}
\begin{center}
\includegraphics*[angle=0, width=0.5\textwidth]{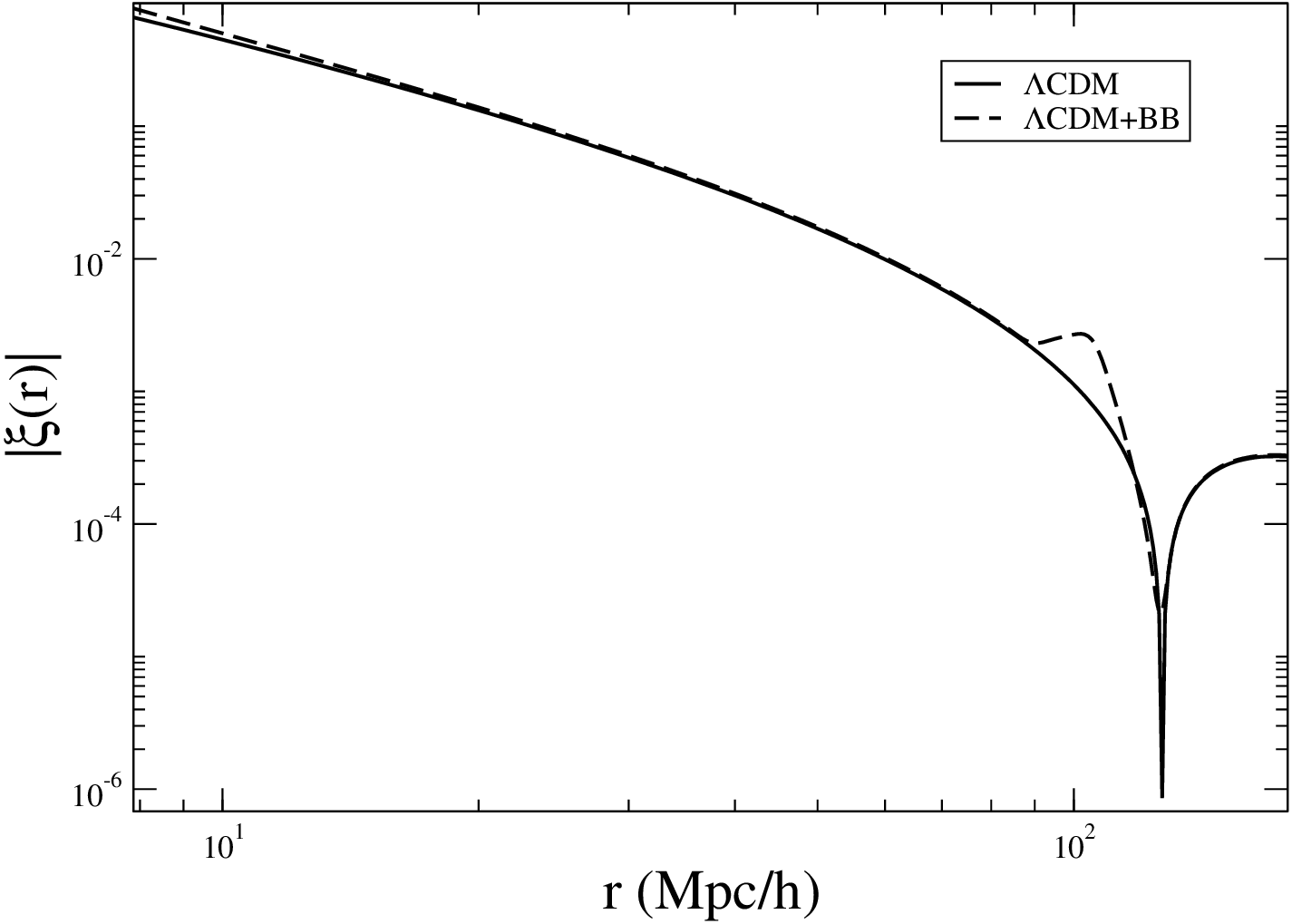} 
\end{center}
\caption{Absolute value 
of the two-point correlation function for the $\Lambda$CDM model
(Eq.\ref{k1}) and for the same model with the baryonic bump (BB) at $\sim $ 100 Mpc/h. }
\label{figbb}
\end{figure}
As one may notice from this  figure the bump appears as a very
small amplitude feature of the two-point correlation function
localized at about $\sim 100$ Mpc/h, i.e. when the correlation
function shows the sharp break corresponding to the approaching to the
zero point, which fixes the global shape of the correlation function
at those scales.  As we discuss below, one of the main problems in the
estimation of the correlation function at such scales in a given
finite sample is to establish whether the break of the power-law
behavior, that is the overall shape corresponding to the presence of
the zero-point scale, is biased or not by a finite size effect. Once
one can be sure enough that the shape is not affected by systematic
effects, then one may try to characterize the presence of the baryonic
feature.


\section{Estimation of the correlation function}

Different estimators of the two-point correlation function have been
introduced and discussed in the literature. The difference between
them lies in their respective method of edge corrections (Kerscher,
Szapudi and Szalay, 2000) which gives rise to different variance and
systematic effects or biases. We discuss three of them (i) the
full-shell (FS) estimator (Gabrielli et al., 2004), (ii) the Landy and
Szalay (LS) estimator (Landy and Szalay, 1993) and (iii) the Davis and
Peebles (DP) estimator (Davis and Peebles, 1983).  The first one has the
advantage that all biases can be carefully understood and possibly
taken under control.  The second is very popular because it has the
minimal variance for the case of a Poisson distribution, although it
has not been demonstrated that the same minimal variance applies in
case of correlated distributions (see e.g. Kerscher, Szapudi and
Szalay, 2000).  However it has the disadvantage that the biases are
very poorly understood in the general case as in the case of the DP
estimator. Although there have been several studies of these
estimators (see e.g. Kerscher, 1999 and Kerscher, Szapudi and Szalay, 
2000) systematic tests for biases are still not completely developed.
Here we give an introduction to the problem and analyze the case of
the FS estimator while in the next section we try to quantify the
problem by studying numerical simulations.

Note that there are, at least, other three estimators known in
the literature, the natural estimator, the Hewett estimator and the
Hamilton estimator which are generally biased as the LS and DP
estimators. In a detailed comparison between these estimators
performed by Kerscher et al. (2000) it is reported that the
performance of the LS estimator is almost indistinguishable from the
Hamilton estimator. In addition Kerscher et al. (2000), after a
careful study, have stressed that LS estimator is the recommended
one. For this reason we decide to focus our studies on the LS while we
have chosen the DP for the reason that it is commonly used in the
literature.

\subsection{Bias in the estimators} 

Let us call $\overline {X(V)}$ the statistical estimator of an average
quantity $\langle X \rangle $ in a volume $V$ (where $\langle X \rangle
$ denotes the ensemble average and $\overline {X}$ the sample
average). In order to be a valid estimator $\overline {X(V)}$ must
satisfy (Gabrielli, et al., 2004)
\be
\label{bias1}
\lim_{V\rightarrow \infty} \overline{X(V)}  = \langle X \rangle \;. 
\ee
A stronger condition is that the ensemble average of the estimator, in
a finite volume $V$, is equal to the ensemble average $\langle X
\rangle$:
\be
\label{bias2} 
\langle \overline {X(V)} \rangle = \langle X \rangle \;. 
\ee
An estimator is called unbiased if this condition is satisfied,
otherwise there is a systematic bias in the finite volume relative to
the ensemble average.  Any estimator $\overline{ \xi(r)}$ of the
correlation function $\xi(r)$, is generally biased. This is because of
the fact that the estimation of the sample mean density is biased when
correlations extend over the sample size and beyond. In fact the most
common estimator of the average density is
\be
\label{sd} 
\overline n = \frac{N}{V} \;, 
\ee
where $N$ is the number of points in a sample of volume $V$.
It is simple to show that (see, e.g., Gabrielli et al., 2004)
\be
\label{biasave}
\langle \overline n \rangle = \langle n \rangle \left( 1 +
\frac{1}{V} \int_V d^3 r \xi(r) \right) \;. 
\ee
Therefore only in case when $\xi(r) =0$ (i.e. for a Poisson
distribution) Eq.\ref{sd} is an unbiased estimator of the ensemble
average density.

In Kerscher (1999) one may find a detailed treatment of estimators of
the two-point correlation function: it has been shown that in a given
sample, on large scales, the biases in the above mentioned estimators
are not negligible especially when there are structures of large
spatial extension inside a given sample. In a $\Lambda$CDM models
there are structures of large amplitude at small scales, i.e. up to
$\sim 10$ Mpc/h, and structures of large spatial extension and low
amplitude up to $\sim$ 120 Mpc/h. Beyond such a scale there will be no
structures anymore as the distribution becomes anti-correlated. Thus it
is important to understand the problem of biases in relation to real
sample estimations, which may cover a distance scale of only several
hundreds Mpc/h, i.e. up to about five times the regime of positive
correlations.

An analytical treatment of the problem, for the general case, is
unfeasible and thus the most direct way to study biases in the
estimators is by performing tests on artificial distributions, which
we discuss in the next section. In what follows we present several
examples which show the importance of the systematic effect related to
Eq.\ref{biasave}, i.e. the fact that the estimators do not satisfy, in
general, Eq.\ref{bias2} but only Eq.\ref{bias1}.

\subsection{The full shell estimator}

The correlation function can be written as 
\be
\xi(r) \equiv \frac{\langle n(r)n(0)\rangle} {n_0^2} -1 \equiv \frac{\cd}{n_0}-1 \;, 
\ee
where the conditional density $\cd = \langle n(r)n(0)\rangle/n_0$ gives the average number of
points in a shell of radius $r$ and thickness $dr$ from an occupied
point of the distribution. Thus FS estimator (Gabrielli et al., 2004)
can be simply written as
\be
\label{xifs1}
\overline{ \xi(r)}  = \frac{\overline{ (n(r))_p}}{\overline n} -1 \,, 
\ee
where $\overline n$ is the estimated number density in the sample and
$\overline{ (n(r))_p}$ is the estimator of the conditional
density. The latter can be written as
\be
\label{cond}
\overline{ (n(r))_p}= 
\frac{1}{N_c(r)} \sum_{i=1}^{N_c(r)} \frac{\Delta N_i (r, \Delta r) }{\Delta V} \;, 
\ee
where $\Delta N_i(r, \Delta r)$ is the number of points in the shell of radius $r$,
thickness $\Delta r$ and volume $\Delta V = 4 \pi r^2 \Delta r$
centered on the $i^{th}$ point of the distribution. Note that the
number of points $N_c(r)$ contributing to the average in Eq.\ref{cond}
is scale dependent, as there are considered only those points such that
when chosen as a center of the sphere of radius $r$, this is fully
included in the sample volume (see Gabrielli, et al., 2004, Vasilyev,
Baryshev, Sylos Labini 2006, for more details). 

The sample density can be estimated in various ways. Suppose that the
sample geometry is simply a sphere of radius $R_s$. The most
convenient in this context is to choose
\be
\label{ne1}
\overline n= \frac{3}{4\pi R_s^3} \int_0^{R_s}  \overline{ (n(r))_p} 4\pi r^2 dr \;,
\ee
as in this case the following integral constraint is satisfied 
\be
\label{xifs}
\int_0^{R_s} \overline{ \xi(r)} r^2 dr = 0 \;.
\ee
This condition is {\it satisfied independently on the functional shape
of the underlying correlation function $\xi(r)$.}

The scale $R_s$, for a sample of arbitrary geometry, is given by the
radius of the maximum sphere fully contained in the sample volume for
the reasons explained above. Other choices for the estimation of the
sample density are possible and give rise to a condition of the type
Eq.\ref{xifs}, even if not precisely the same.  This condition
introduces a systematic distortion in the measured shape of $
\overline{ \xi(r)}$ and the advantage in choosing Eq.\ref{ne1} lies in
the fact that one has a certain control on the scale $r_*$ defined to
be the scale beyond which the distortion becomes important. The scale
$r_*$ must be evaluated given a specific model for $\xi(r)$, but it is
in general a fraction of $R_s$.

Thus the integral constraint for the FS estimator, Eq.\ref{xifs}, does
not simply introduce an offset, but a change in the functional
behavior of the estimated correlation function. Other choices
introduce distortions at a scale which is difficult to be evaluated
especially in the case the sample does not have a simple spherical
geometry. In general any estimator is distorted at some scales by a
condition of the type given by Eq.\ref{xifs}, which basically reflects
our ignorance on the value of the ensemble average density.
 
 In order to study the effect of the integral
constraint for the FS estimator, let us rewrite the estimation of the
correlation in terms of the theoretical correlation function
\be
\label{estth}
 \overline{ \xi(r)} = \frac{1+\xi(r)}{1+\frac{3}{R_s^3}\int_0^{R_s}
 \xi(r) r^2 dr} -1 \;. 
\ee
By writing Eq.\ref{estth} we assume that the stochastic noise is
negligible, which of course is not a good approximation at any
scale. However in this way we may be able to understand the effect of
the integral constraint for the FS estimator. From Eq.\ref{estth} it
is clear that this estimator is biased, as it does not satisfy
Eq.\ref{bias2} but only Eq.\ref{bias1}.

Let us consider two useful examples for the theoretical correlation
function (i) $\xi(r) \sim r^{-\gamma}$ and in (ii) $\Lambda$CDM model
of Eq.\ref{xi2}. The distortion due to the integral constraint in the
FS estimator in the case the theoretical correlation function has a
power-law behavior with exponent $\gamma=2$ is illustrated in
Fig.\ref{figicpl2}. One may see that at $r \approx R_s/3$ the
estimation is already distorted and when $r\approx R_s/2$ the function 
$\overline{ \xi(r)} $ crosses zero and becomes negative in order to
satisfy Eq.\ref{xifs}.
\begin{figure}
\begin{center}
\includegraphics*[angle=0, width=0.5\textwidth]{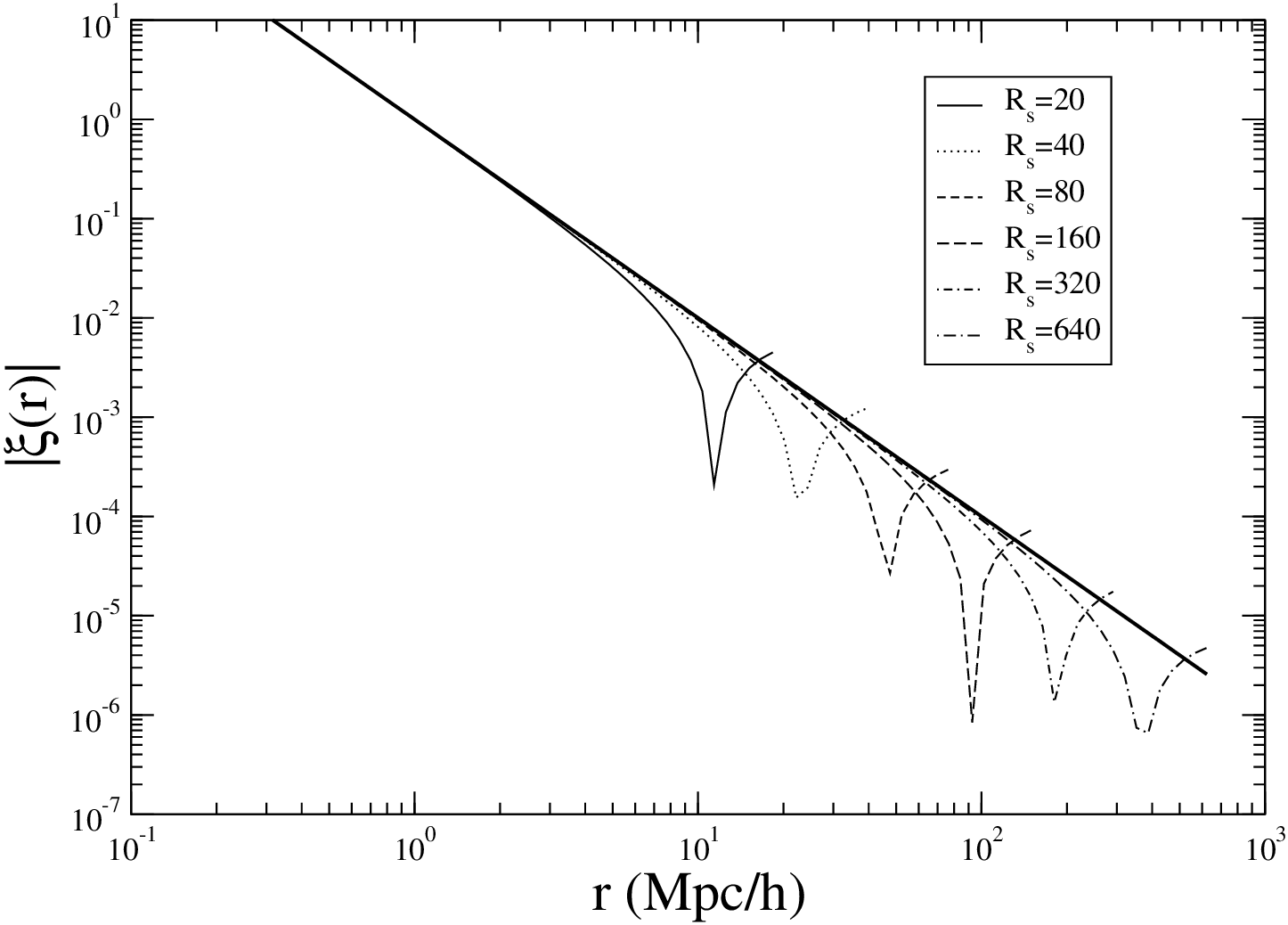} 
\end{center}
\caption{Absolute value of the estimation of the correlation
 function $\xi(r)\sim r^{-\gamma}$, with $\gamma=2$, by the FS
 estimator. The tick solid line represents the theoretical model. 
The condition given by the integral constraint described
 by Eq.\ref{estth} is taken into account: beyond the scale at which
 there is the break of the power-law behavior the correlation function
 crosses zero and becomes negative.}
\label{figicpl2}
\end{figure}

The case of the $\Lambda$CDM model is shown in Fig.\ref{figicvanilla}.
The situation is similar to the power-law case as long as one
considers $R_s$ smaller than the zero point scale $r_c$.  For larger
$R_s$ one may see that zero point is not changed anymore, while the
negative tail continues to be amplified in a non-linear way even at
scales $r<R_s$. For example with a sample of size $R_s\approx 600$
Mpc/h the distortion of the power-law tail does not allow to detect
the $\xi(r)
\sim - r^{-4}$ behavior which is marginally visible only when $R_s >
1000$ Mpc/h.
\begin{figure}
\begin{center}
\includegraphics*[angle=0, width=0.5\textwidth]{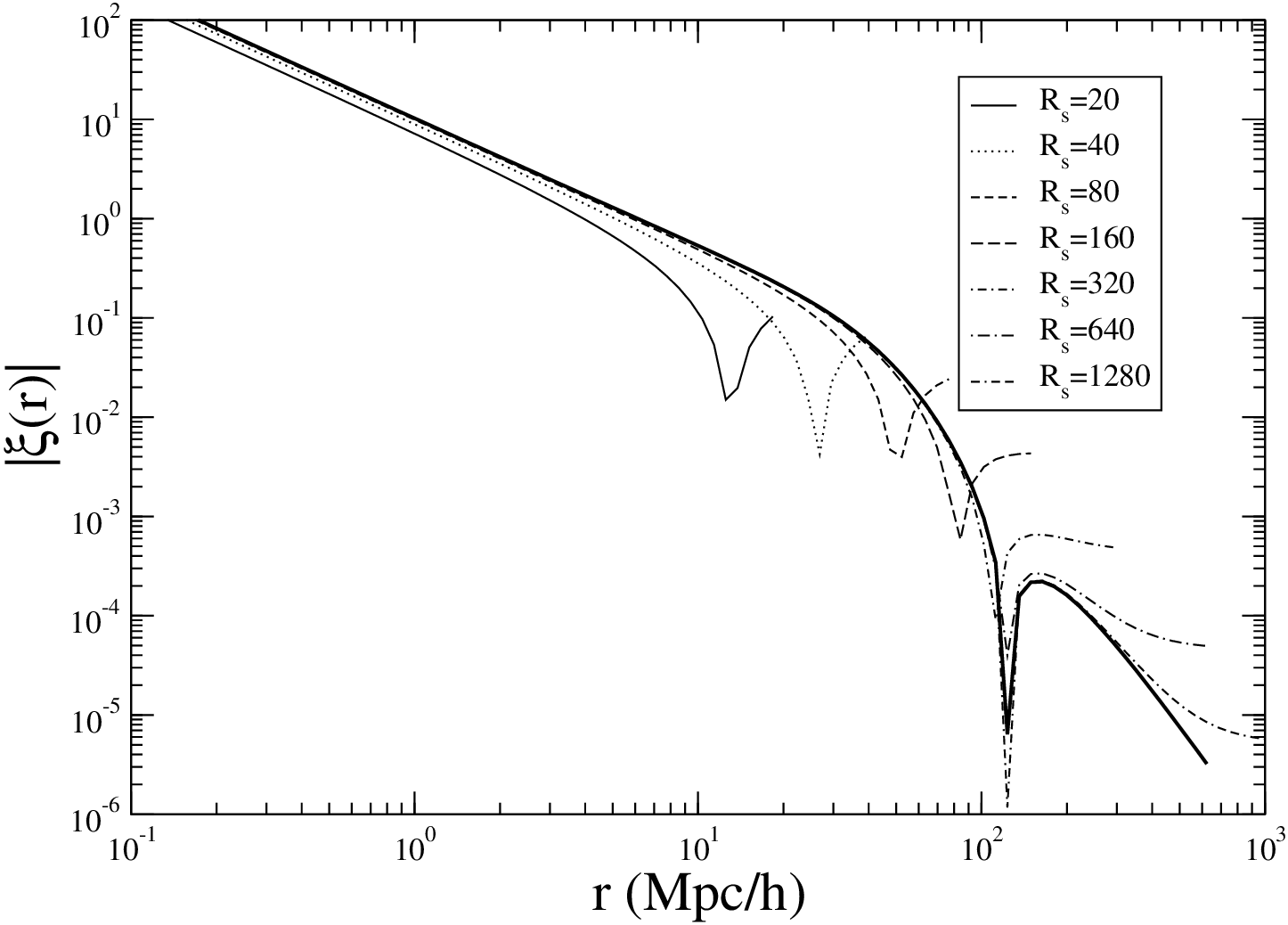} 
\end{center}
\caption{Absolute value of the estimation of the correlation function  of the  $\Lambda$CDM  model 
with the integral constraint described by Eq.\ref{estth}. The tick
solid line represents the theoretical model.}
\label{figicvanilla}
\end{figure}


\subsection{Pairwise estimators}

To determine a pairwise estimator we define the following
quantities. The number of data-data pairs
\be
\label{pw1}
DD(r) = \sum_{i}^{N_d} dd_i(r,\Delta r) \;,
\ee
the number of data-random pairs
\be
\label{pw2}
DR(r) = \sum_{i}^{N_d} dr_i(r,\Delta r) \;,
\ee
the number of random-random pairs
\be
\label{pw3}
RR(r) = \sum_{i}^{N_r} rr_i(r,\Delta r) \;.
\ee
where $N_d$ is the number of data points, $N_r$ is the number of
random points, which are Poisson distributed, $dd_i(r,\Delta r)$,
$dr_i(r,\Delta r)$ and $rr_i(r,\Delta r)$ are respectively the numbers
of data-data, data-random and random-random pairs in the shell of
radius $r$ and thickness $\Delta r$ around the $i^{th}$ center.

The DP estimator is defined as (Davis and Peebles, 1983) 
\footnote{For seek of clarity hereafter we denote the {\it estimator} 
as $\xi_{XX}$ where XX can be FS for the full-shell case, 
DP for the Davis and Peebles case and LS for the Landy and Szalay case.
We omit the $\overline{X}$ symbol which was previously 
introduce to mean that this is an estimator of the 
statistical quantity X.} 
\be
\label{dp}
\xi_{DP} (r) = \frac{N_r}{N_d-1} \frac{DD(r)}{DR(r)} -1 \;.
\ee

The LS estimator is defined as (Landy and Szalay, 1993) 
\be
\label{ls}
\xi_{LS} (r) = \frac{N_r(N_r-1)}{N_d(N_d-1)} \frac{DD(r)}{RR(r)} - 
2\frac{N_r-1}{N_d} \frac{DR(r)}{RR(r)} +1  \;. 
\ee

Finally the Hamilton estimator is defined as (Hamilton, 1993)
\be
\label{hamilton}
\xi_{H} (r) = \frac{N_rN_d}{(N_r-1)(N_d-1)} \frac{DD(r)RR(r)}{DR^2(r)} -1  
\;. 
\ee

\subsection{Errors}
The determination of measurement errors of the correlation
function can be performed in various ways.  This first is a
calculation of the error on $\overline{ \xi(r)} $ in a given sample using
the Poisson estimate (Ross et al., 2007)
\be
\label{err_poiss}
\sigma^2_P (r) = \frac{1+ \overline{ \xi(r)} }{\sqrt{DD(r)}} \;. 
\ee
The second error estimation method is the field-to-field error, which 
is obtained by divining the whole sample into $N$ subsamples and by computing
in each of these the correlation function $\overline{ \xi_i(r)} $ for $i=1...N$
\be
\label{err_ftf}
\sigma^2_{FtF} (r)= \frac{1}{N-1} \sum_{i=1}^{N} 
\frac{DR_i(r)}{DR(r)} \left( \overline{ \xi_i(r)}  - \overline{ \xi(r)} \right)^2 \;,
\ee
and $\overline{ \xi(r)} $ is the estimation of the correlation function
in the whole sample.  The third method is called jackknife estimate
(Scranton et al., 2002, Zehavi et al., 2004) and the variance is
estimated by
\be
\label{jackerrors}
\sigma^2_{Jack} (r)= \sum_{i'=1}^{N} \frac{DR_{i'}(r)}
{DR(r)} \left( \overline{ \xi_{i'}(r)} - \overline{ \xi(r)} \right)^2
\ee
where the index $i'$ is used to signify that each time the value of
the correlation function $\overline{ \xi_{i'}(r)} $ is computed in all
subsamples but one (the $i^{th}$). Finally another possibility is to
divide the sample into $N$ subfields, to compute the average
\be
\label{xiave}
\overline{ \xi(r)}  = \frac{1}{N} \sum_{i=1}^{N} \xi_i(r)
\ee
and then the variance on the average 
\be
\label{xisigma} 
\sigma_{a}^2 (r) = 
\frac{1}{N} \sum_{i=1}^{N} \frac{(\xi_i(r) - \overline{ \xi(r)} )^2}{N-1} \;.
\ee
We show in what follows that Eq.\ref{xisigma} is equivalent
to Eq.\ref{jackerrors} at all but the largest scales of a the sample
where it gives a more conservative estimation of the errors. 
In what follows we will make use of the errors estimated by Eq.\ref{xisigma}
which are similar to the jackknife ones (Eq.\ref{jackerrors}). Below
we discuss in details the determination of the errors in 
artificial distributions and a comparison between the different
methods to define them.

\section{Test on artificial distributions}

We consider a distribution of points extracted from a cosmological
N-body simulation generated in framework of the Millennium project
(Springel, et al. 2005), which consists of $N=6,528,040$ particles in
a cubic box of nominal side $L=1$ and which is one of the
semi-analytic catalogs (Croton et al., 2006) constructed to produce
mock galaxy samples. This distribution presents strong clustering up
to a scale of $r_0 \approx 0.01$ and then it presents weak power-law
correlations up to the sample size. We compare the results of each
estimator in the sub-boxes of varying size with the determination of
the FS estimator in the box of side $L=1$ which we take as a
reference.  In principle, one would like to have a theoretical
prediction to compare with: however due to the effect of the formation
of non-linearities and to the sampling used to produce these
distributions, one does not have a simple way to compute the
theoretical correlation function. This is the reason why we have
chosen the correlation function computed in the entire box as a
reference. In addition, for all statistical quantities considered, we
limit our analysis to the scale $R_s=0.2$ in order to minimize finite
size effects. In what follows we report the results by using the
field-to-field average quantities and variance (i.e.  Eq.\ref{xiave}
and Eq.\ref{xisigma}) which we find to be the most conservative error
determinations.  Below we also present a discussion of the different
determinations of the errors.

\subsection{Cubic Samples} 

We have divided the box of side $L=1$ into $N_f$ non overlapping
sub-boxes of side $\ell =0.05,0.1,0.15,0.2,0.25$ and we have computed
the correlation function in each of the sub-boxes. Note that the
number sub-boxes over which the calculations are performed is taken to
be constant independently on their size and $N_f=16$.  The
determination of the correlation function by using the FS estimator is
shown in Fig.\ref{figxifs}.  The main difference between the estimated
correlation function and the ``true'' one is due to the integral
constraint. This can be shown by the comparison of the estimated
correlation function with that computed by using Eq.\ref{estth} which
describes the effect of the integral constraint.
\begin{figure}
\begin{center}
\includegraphics*[angle=0, width=0.5\textwidth]{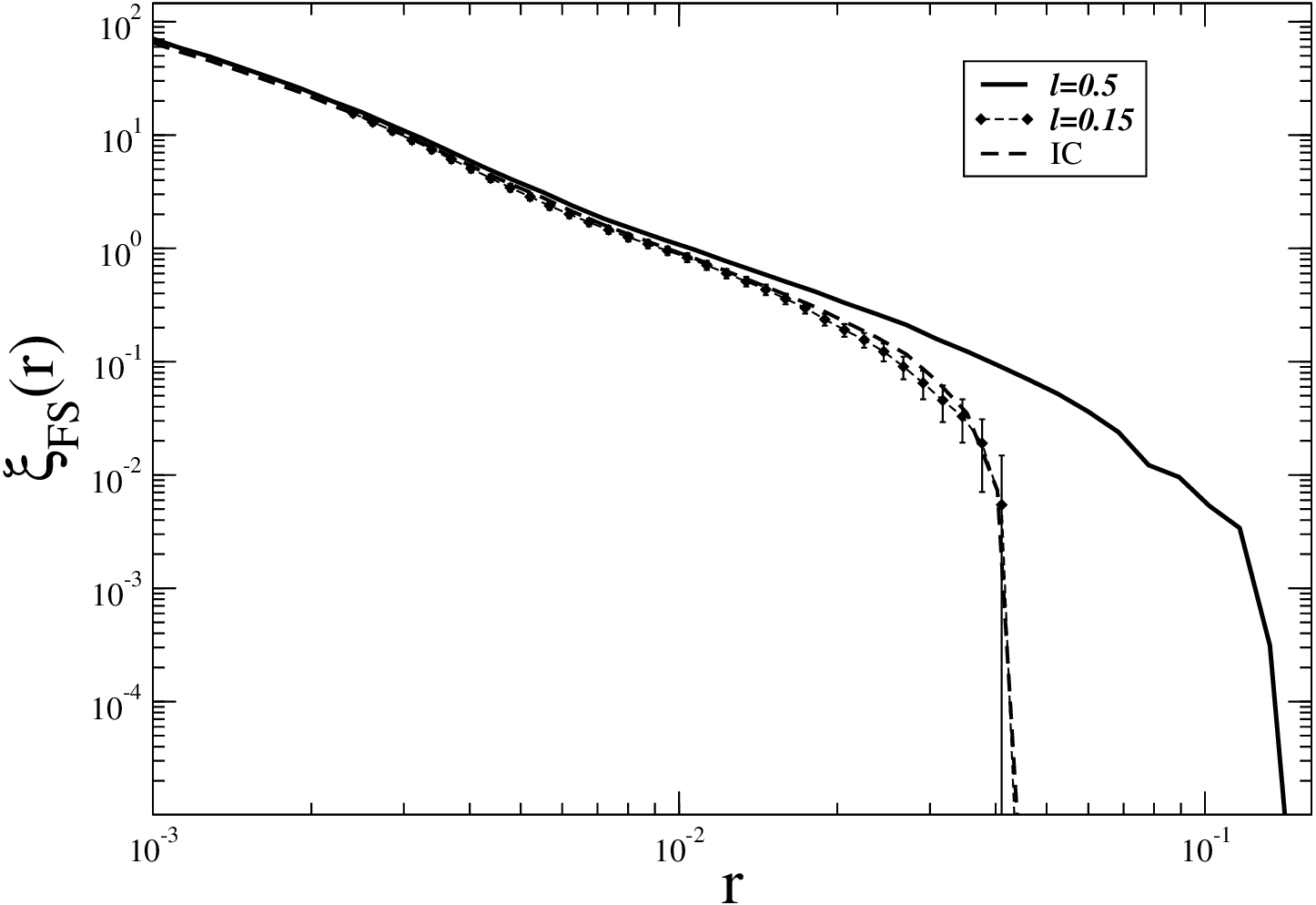} 
\end{center}
\caption{Average 
correlation function computed by using the FS estimator in independent
sub-boxes of side $\ell=0.15$ together with the prediction of
Eq.\ref{estth}. The solid line represents the correlation function
computed by using the FS estimator in independent sub-boxes of side
$\ell=0.5$ while the dotted line (IC) represents the 
analytical computation of the estimated correlation 
with the integral constraint (i.e. Eq.\ref{estth}). }
\label{figxifs}
\end{figure}

In Fig.\ref{figxifs3} we compare the determinations of the correlation
function by the FS estimator in sub-boxes of different sizes.
\begin{figure}
\begin{center}
\includegraphics*[angle=0, width=0.5\textwidth]{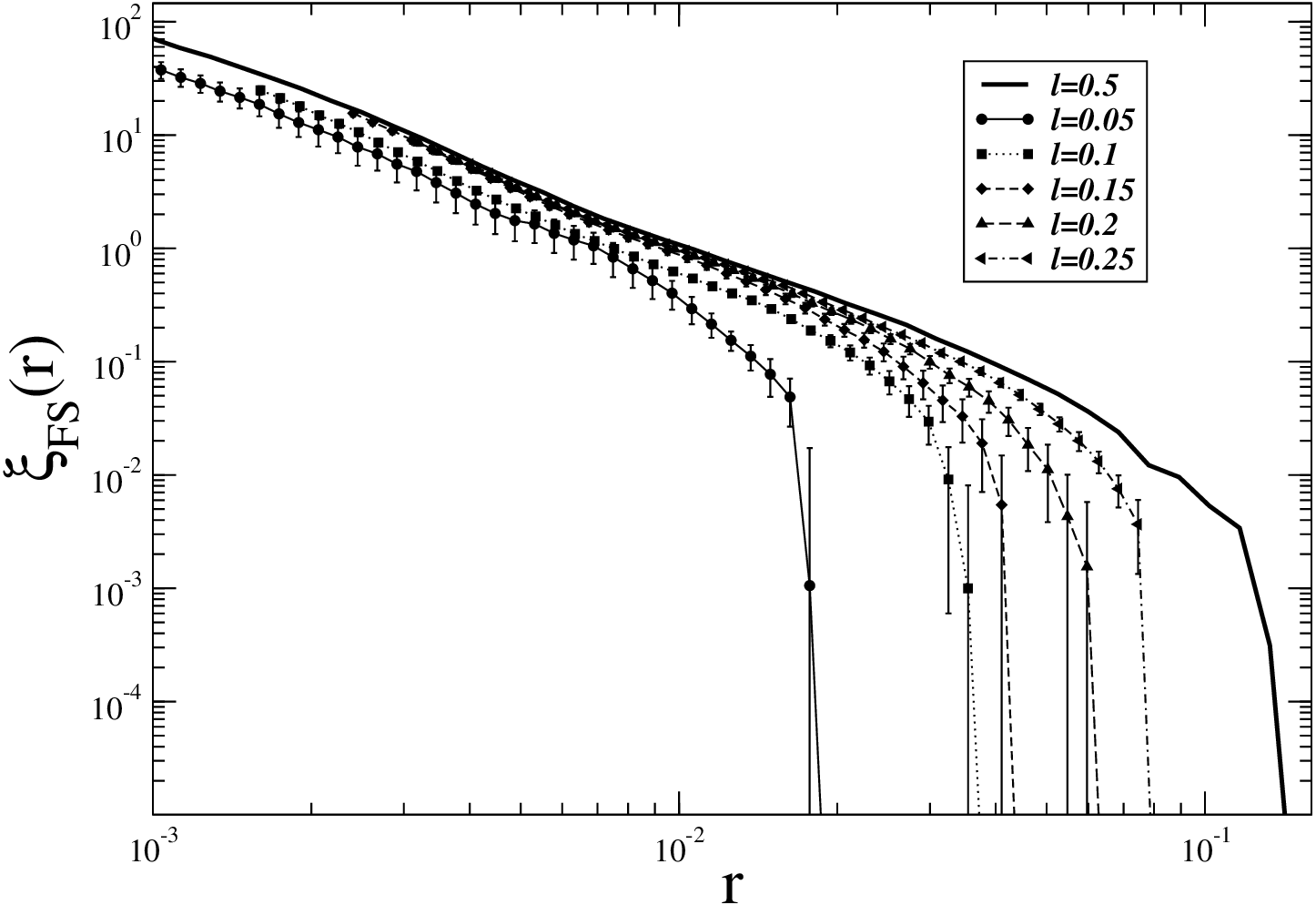} 
\end{center}
\caption{Average 
correlation function computed by using the FS estimator in independent
sub-boxes of side $\ell=0.05,0.1,0.15,0.2,0.25$ respectively. The
solid line represents the correlation function computed by using the
FS estimator in independent sub-boxes of side $\ell=0.5$.}
\label{figxifs3}
\end{figure}
One may note that the effect of the integral constraint, for what
concerns the amplitude of the estimated correlation function, is
important for the subsamples with $\ell \le 0.1$ as in this case the
distribution is strongly non-linear inside the sample thus the
determination of the sample density strongly depends on the sample
size. This is shown by both a smaller amplitude and a smaller range of
distance scales over which the correlation function is positive. The
break in the positive behavior occurs at a distance scale of order
$\ell$ independently on the amplitude of the correlation
function. This is again a finite-size effect which can be easily
understood as due to the integral constraint.

To summarize there are two distinct effects: (i) the amplitude of the
estimated correlation function strongly depends on the sample size
when the distribution exhibits strong clustering and (ii) the
artificial break of the positive correlations is sample-size dependent.

In Figs.\ref{xifs5}-\ref{xifs9} we compare the FS, DP and LS
estimators. One may note that the DP and LS estimators are biased by a
similar effect as the FS estimator, due to the integral constraint,
although the break in the power-law behavior seems to occur at
slightly larger scales than for the FS estimator. This difference can
be attributed to the fact that the LS and DP estimators implicitly use
the estimations of the average density at scale $\ell$ instead of at
the scale $\ell/2$ as the FS estimator.  To clarify this point in the
next section we present some other tests which have been tuned to
explore this effect. 

\begin{figure}
\begin{center}
\includegraphics*[angle=0, width=0.5\textwidth]{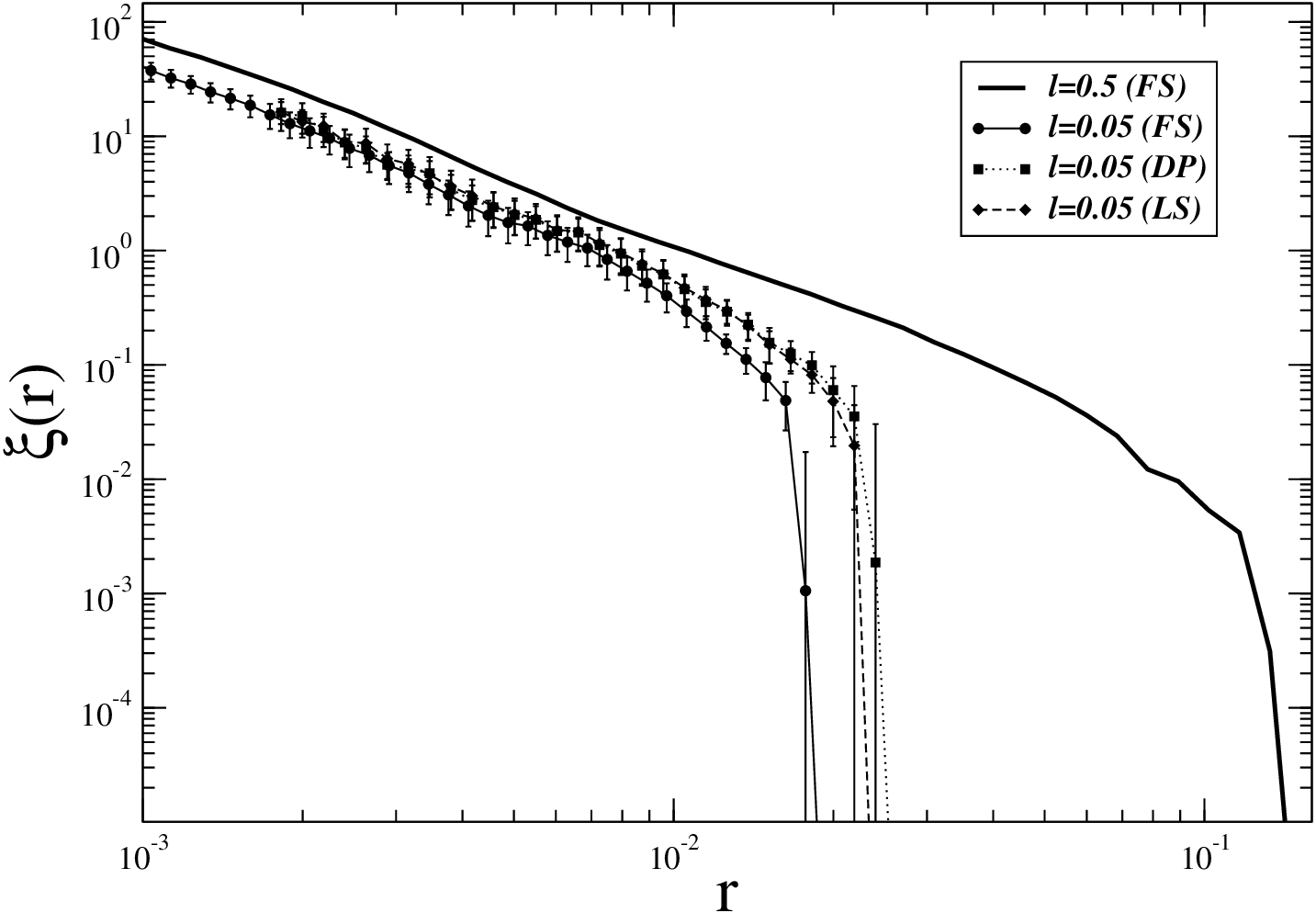}
\end{center}
\caption{Average 
correlation function by using the FS, LS and DP estimator respectively
in independent sub-boxes of side $\ell=0.05$.  The solid line represents the correlation function
computed by using the FS estimator in independent sub-boxes of side
$\ell=0.5$.}
\label{xifs5}
\end{figure}

\begin{figure}
\begin{center}
\includegraphics*[angle=0, width=0.5\textwidth]{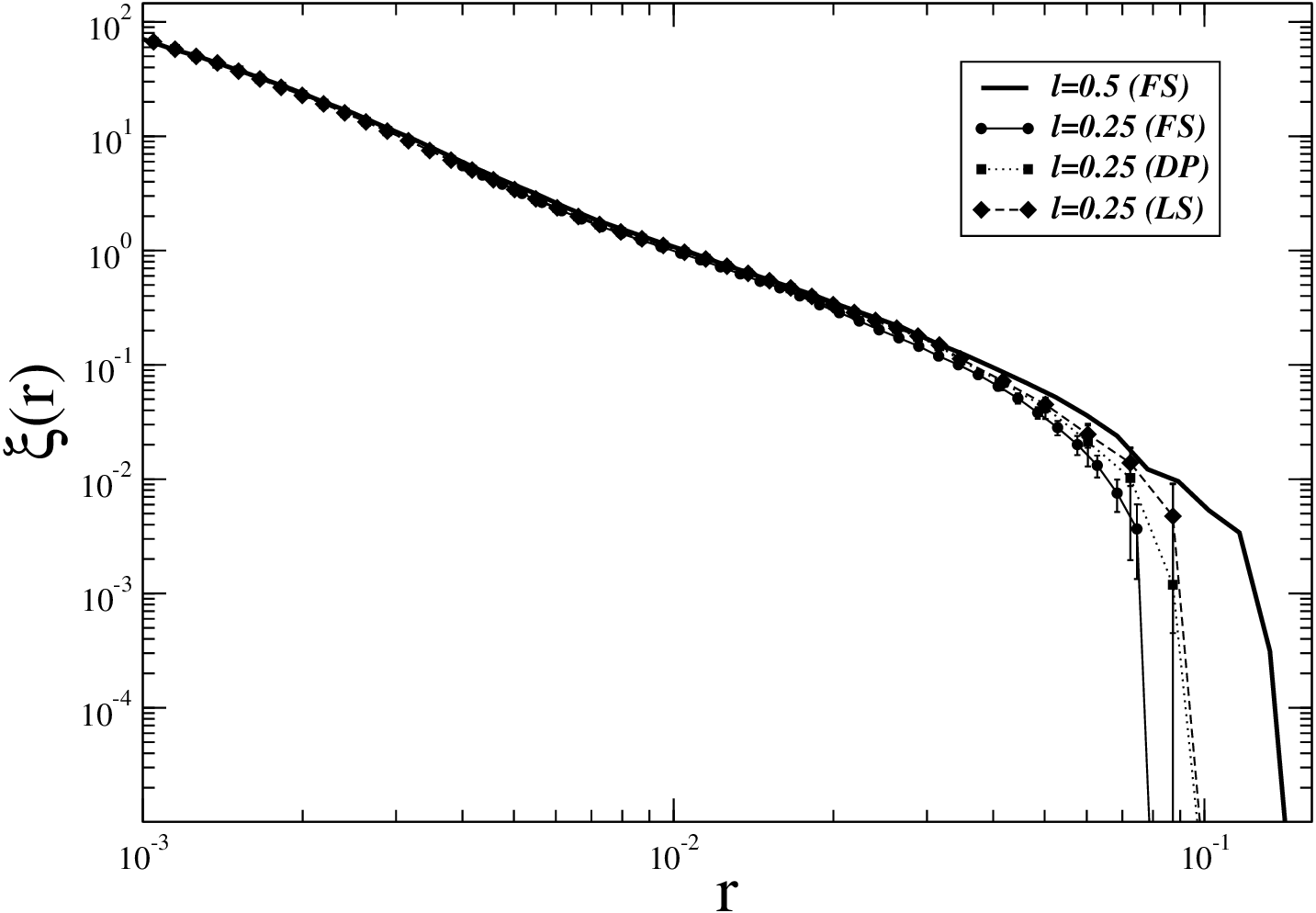}
\end{center}
\caption{Average 
correlation function computed by using the FS, LS and DP estimator respectively
in independent sub-boxes of side $\ell=0.25$. The solid line represents the correlation function
computed by using the FS estimator in independent sub-boxes of side
$\ell=0.5$.}
\label{xifs9}
\end{figure}

In Fig.\ref{xifs5b} we compare the LS and Hamilton
estimators. We confirm the results of Kerscher et al. (2000) that the
Hamilton and LS estimators give indistinguishable results, inside the
error bars, and thus we will focus on the former hereafter.
\begin{figure}
\begin{center}
\includegraphics*[angle=0, width=0.5\textwidth]{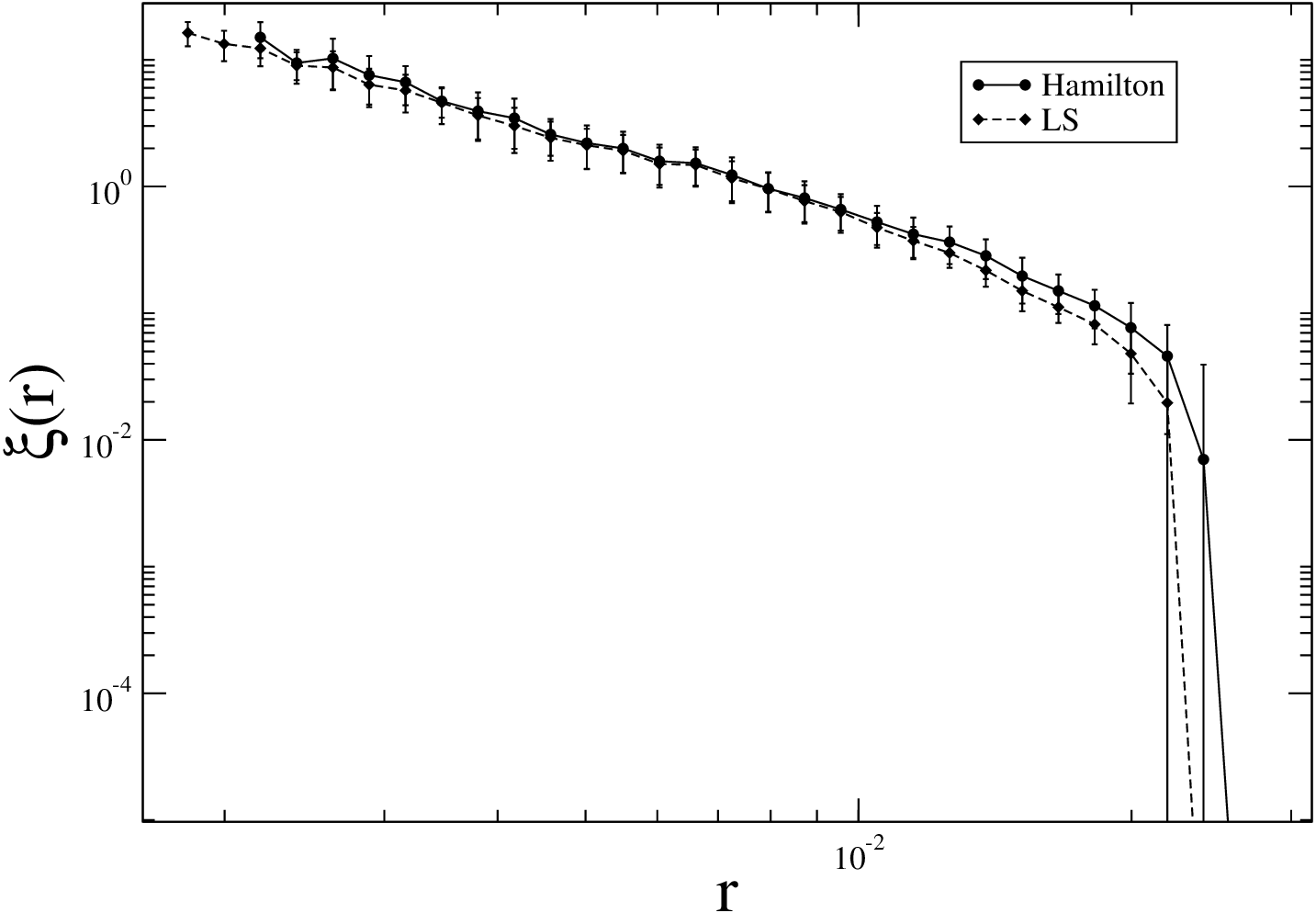}
\end{center}
\caption{
Average 
correlation function by using the LS and Hamilton estimator
respectively in independent sub-boxes of side $\ell=0.05$.}
\label{xifs5b}
\end{figure}

In Figs.\ref{xifs10}-\ref{xifs11} we show the determinations of the
average correlation function computed by using the LS and DP
estimators in independent sub-boxes of side $\ell=$0.05, 0.1, 0.15,
0.2, 0.25 respectively: the finite size dependencies of the amplitude
and of the break are still present as for the FS estimator, and
analogously to this former case, they can be understood as an effect
of the integral constraint.

\begin{figure}
\begin{center}
\includegraphics*[angle=0, width=0.5\textwidth]{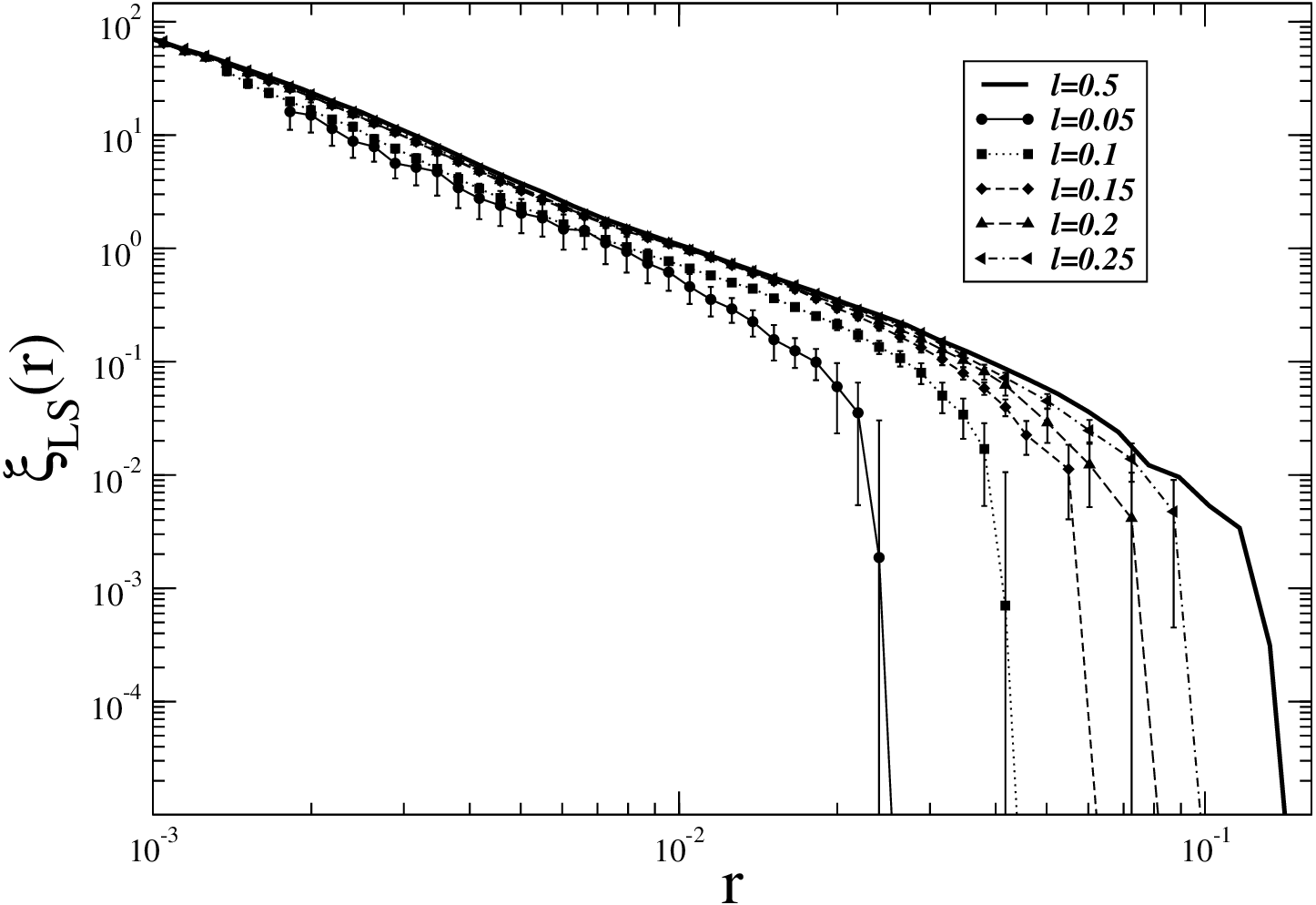}
\end{center}
\caption{
Average 
correlation function computed by using the LS estimator 
in independent sub-boxes of side $\ell=0.05,0.1,0.15,0.2,0.25$ respectively. The solid line represents the correlation function
computed by using the FS estimator in independent sub-boxes of side
$\ell=0.5$.}
\label{xifs10}
\end{figure}
\begin{figure}
\begin{center}
\includegraphics*[angle=0, width=0.5\textwidth]{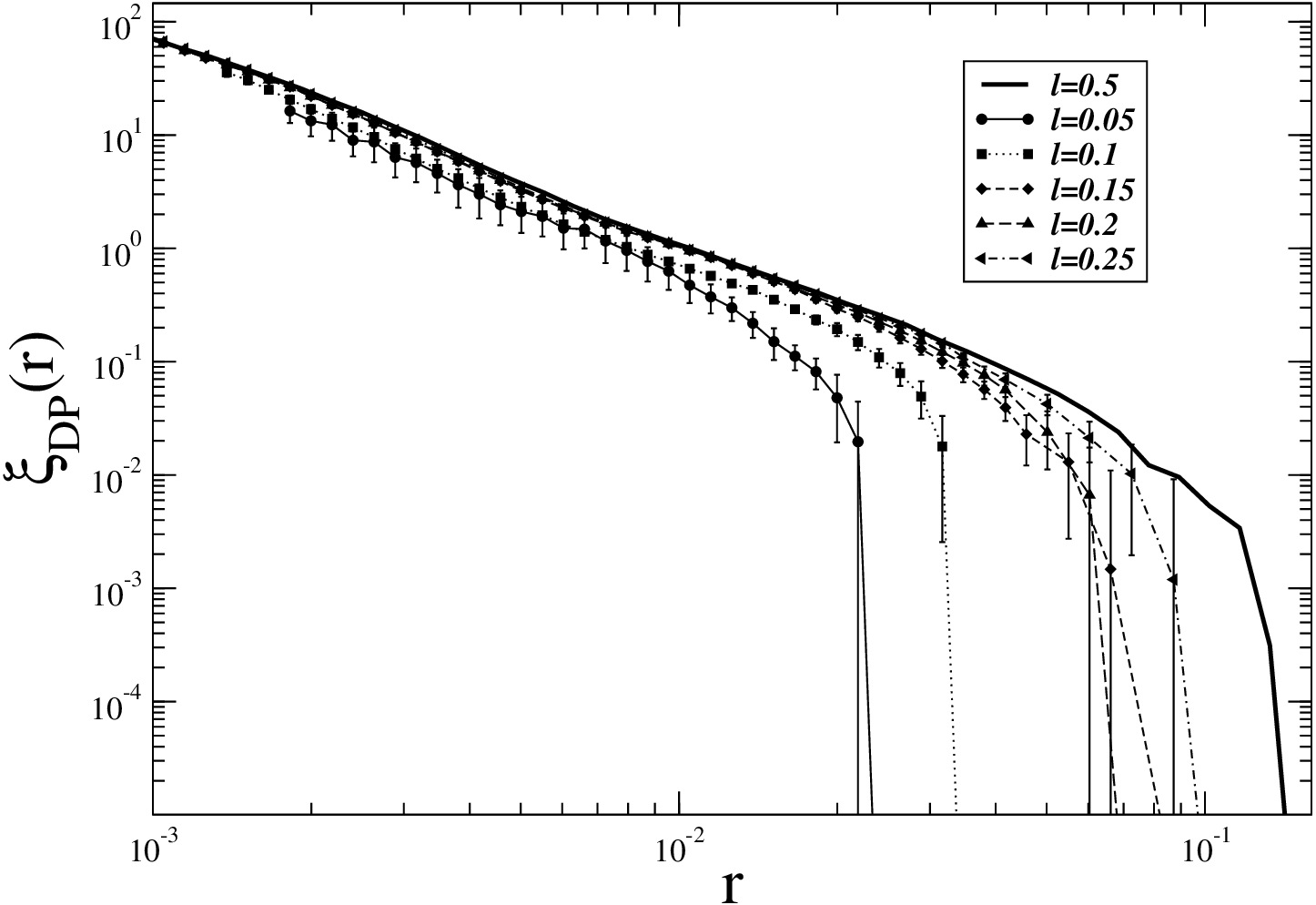}
\end{center}
\caption{Average 
correlation function computed by using the DP estimator 
in independent sub-boxes of side $\ell=0.05,0.1,0.15,0.2,0.25$ respectively.  The solid line represents the correlation function
computed by using the FS estimator in independent sub-boxes of side
$\ell=0.5$.}
\label{xifs11}
\end{figure}


\subsection{Slices}
 
We have seen that the estimation of the correlation function is
affected by a finite size effect which depends on the sample size,
which up to now has been considered to be a simple geometrical shape
as a sphere or a cubic box. In order to investigate a situation closer
to real observations we have constructed several subsamples of the
original distribution in the following way. We have considered the
observer placed in the center of the box (0.5,0.5,0.5) and we have
identified a sphere of radius $0.5$ centered on that point. We have
considered the spherical coordinates $\alpha,\delta,r$ of the
distribution points with respect to such center,  where $0 \le \alpha
\le 2 \pi$, $-\pi/2 \le \delta \le \pi/2$ and $0 \le r \le \le 0.5$. It is now 
possible to construct several subsample which have a certain depth
$R_{depth} \le 0.5$ and specific cuts in $\alpha$ and $\delta$. In general
the solid angle of a portion of a sphere is
\be
\Omega = \Delta \alpha \times \Delta \mu \;, 
\ee
where $ \Delta \alpha = \alpha_2 -\alpha_1$ with $\alpha_1,\alpha_2$
the limits in right ascension delimiting the angular region and
$\Delta \mu = \sin(\delta_2) - \sin(\delta_1)$, with
$\delta_1,\delta_2$ the limits is declination delimiting the angular
region. We have chosen $\Delta \mu =2$, i.e. $\delta_1 = -\pi/2$ and
$\delta_2 = \pi/2$ and $\Delta \alpha =$const. In such a way we have
constructed $N_f$ independent spherical slices with constant solid
angle and same geometry. The number of slices is thus $N_f =
2\pi/\Delta\alpha$: we have taken $N_f \le 30$.  We have then computed
the LS and DP estimators and their field-to-field variance
(Eq.\ref{xisigma}).

In Figs.\ref{xislice1}-\ref{xislice3} we show the average correlation
function computed by using the LS and DP estimators respectively in
$N_f=30$ angular slices with $\Delta \alpha = 0.0063,0.013,0.063$
respectively. One may note that the LS and DP estimator are very
similar although the LS estimator extends to slightly large
scales. The amplitude in this case corresponds to the expectation
value for the FS estimator in a box of side $R_{depth} =0.05$ which is about
ten times larger than the radius of the maximum sphere fully included
in the sample volume.

\begin{figure}
\begin{center}
\includegraphics*[angle=0, width=0.5\textwidth]{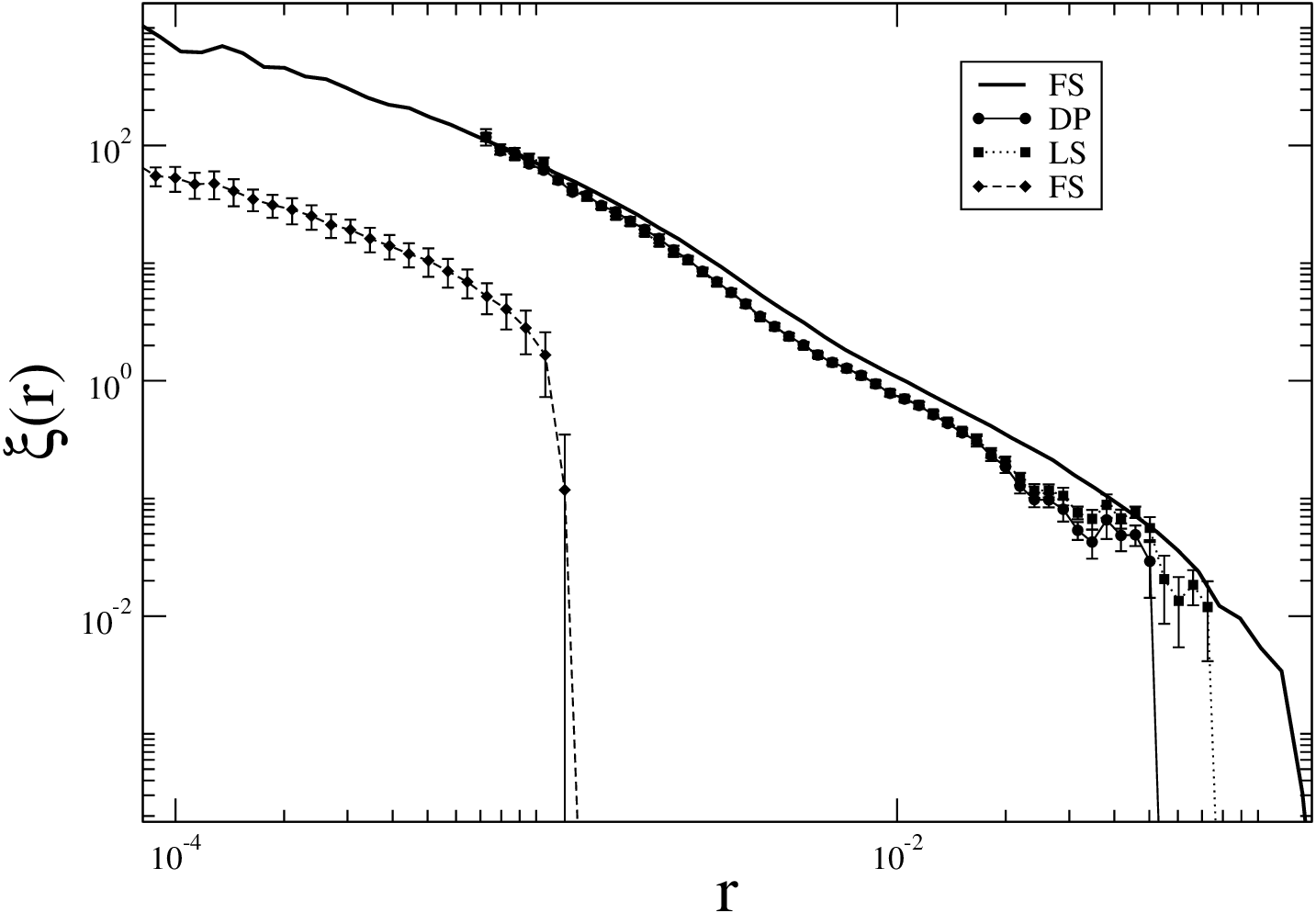} 
\end{center}
\caption{Average 
correlation function computed by using the FS, LS and DP estimator
respectively in $N_f=30$ angular slices with $\Delta \alpha = 0.0063$. 
The solid line represents the correlation function
computed by using the FS estimator in independent sub-boxes of side
$\ell=0.5$. }
\label{xislice1}
\end{figure}

\begin{figure}
\begin{center}
\includegraphics*[angle=0, width=0.5\textwidth]{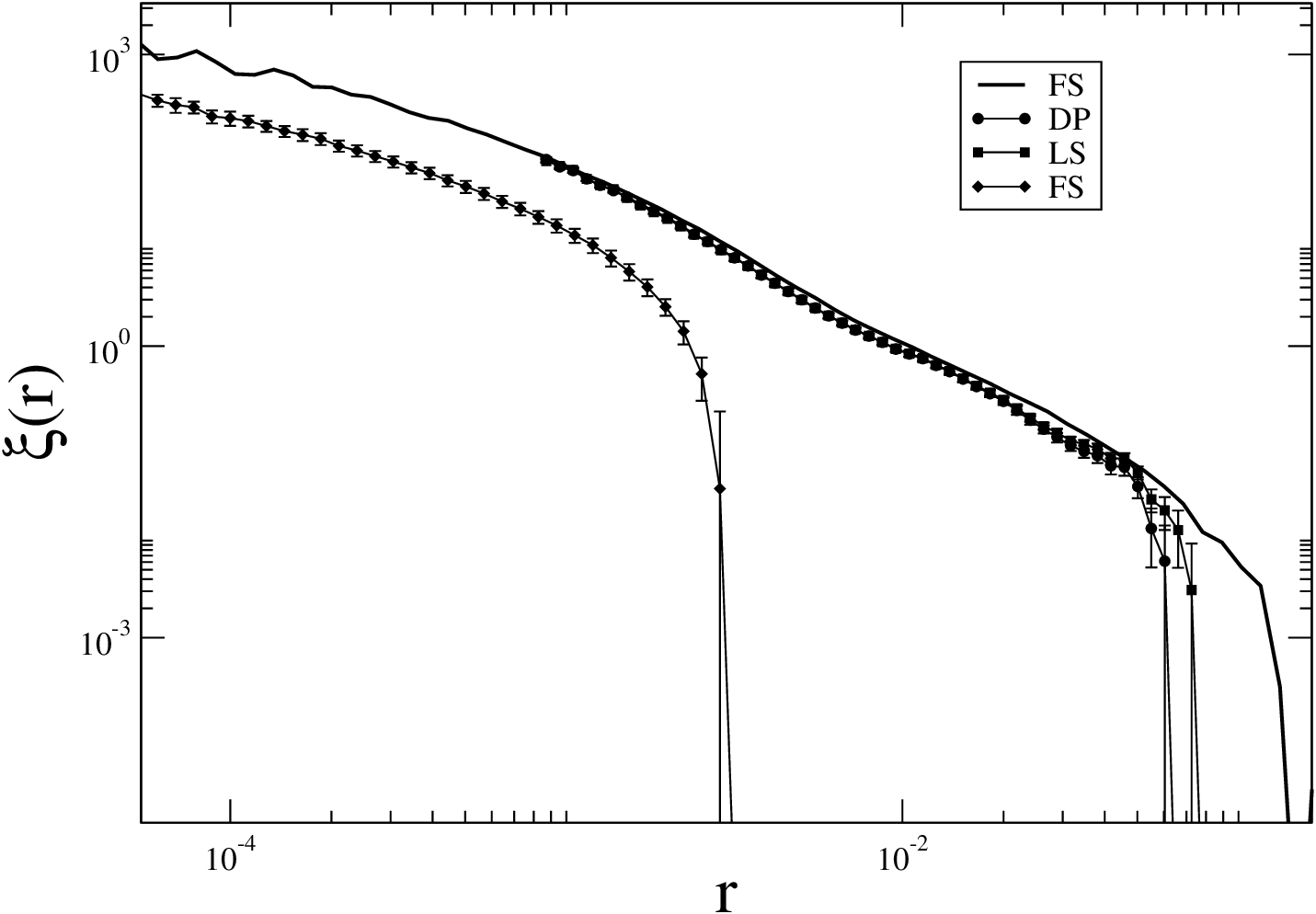}
\end{center}
\caption{Average 
correlation function computed by using the FS, LS and DP estimator
respectively in $N_f=30$ angular slices with $\Delta \alpha = 0.013$. The solid line represents the correlation function
computed by using the FS estimator in independent sub-boxes of side
$\ell=0.5$. }
\label{xislice2}
\end{figure}

\begin{figure}
\begin{center}
\includegraphics*[angle=0, width=0.5\textwidth]{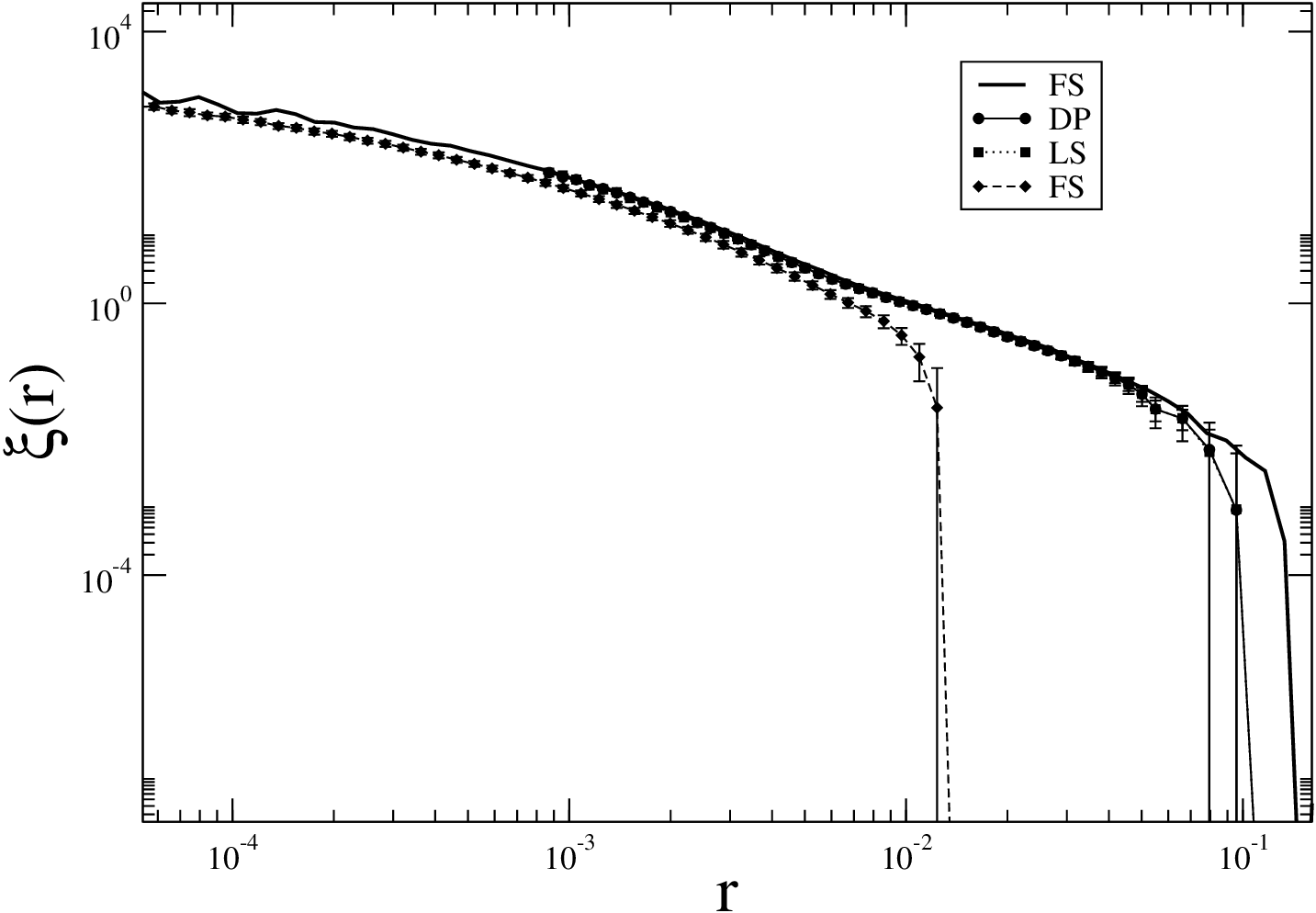}
\end{center}
\caption{Average 
correlation function computed by using the FS, LS and DP estimator
respectively in $N_f=30$ angular slices with $\Delta \alpha = 0.063$. 
The solid line represents the correlation function
computed by using the FS estimator in independent sub-boxes of side
$\ell=0.5$. }
\label{xislice3}
\end{figure}

By comparing (see Figs.\ref{xislice3b}-\ref{xislice3c}) the FS, LS and DP
estimators computed in angular slices with $\Delta \alpha = 0.0063,
0.013, 0.063$ one may note that the amplitude slightly increases by
choosing a larger solid angle and the range of scales where one may
estimate the correlation function also increases when $\Delta \alpha$
increases. The exact location of the break of the power-law behavior
and the value of the amplitude are in agreement with a value of $R_{depth}$
in integral constraint of the order of the sample depth $\ell$ and not
of the radius of the maximum sphere fully enclosed as for the case of
the FS estimator.

\begin{figure}
\begin{center}
\includegraphics*[angle=0, width=0.5\textwidth]{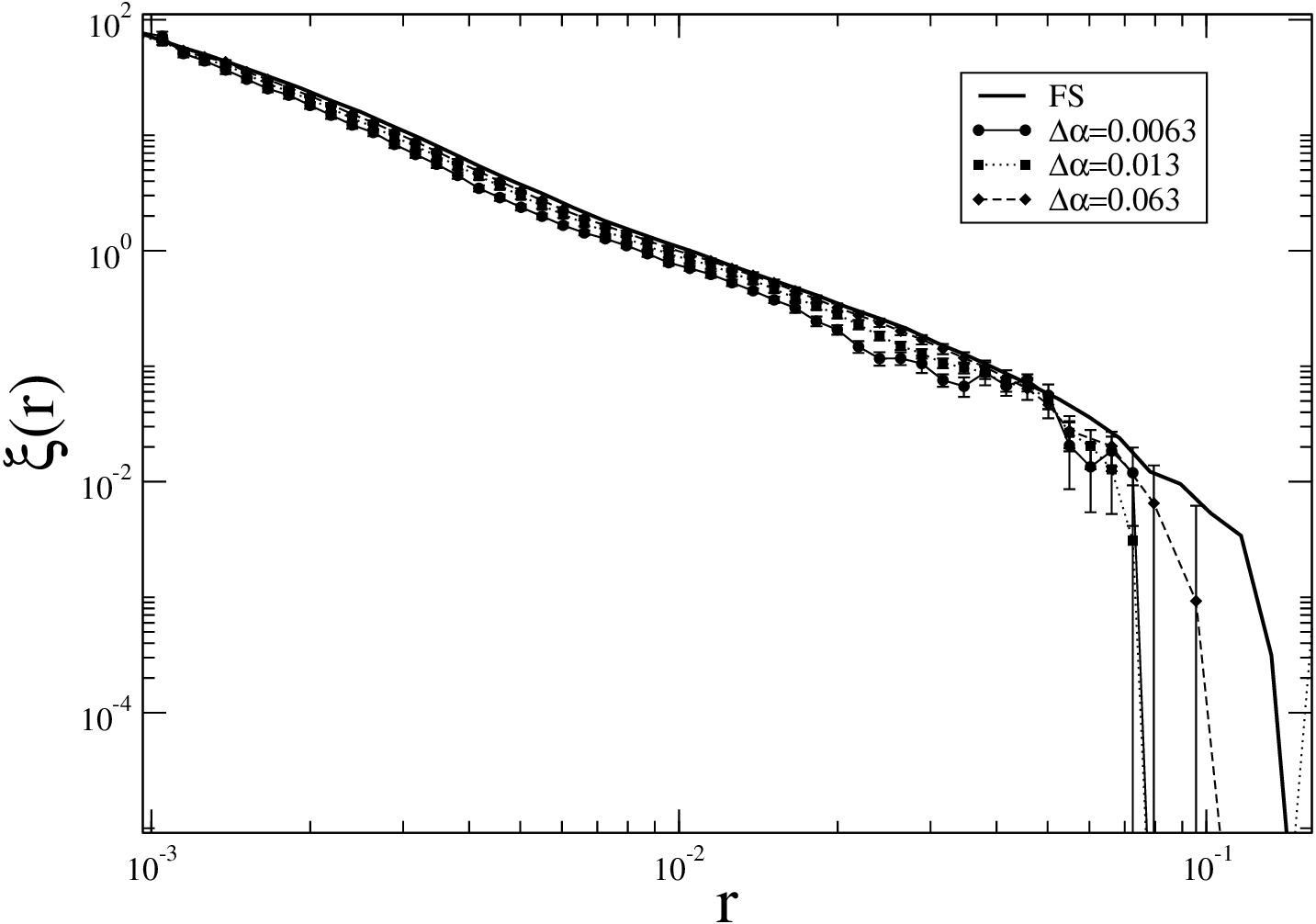} 
\end{center}
\caption{Average 
correlation function computed by using the LS estimator in $N_f=30$
angular slices with $\Delta \alpha = 0.0063, 0.013, 0.063$. The solid
line represents the correlation function computed by using the FS
estimator in independent sub-boxes of side $\ell=0.5$. }
\label{xislice3b}
\end{figure}
\begin{figure}
\begin{center}
\includegraphics*[angle=0, width=0.5\textwidth]{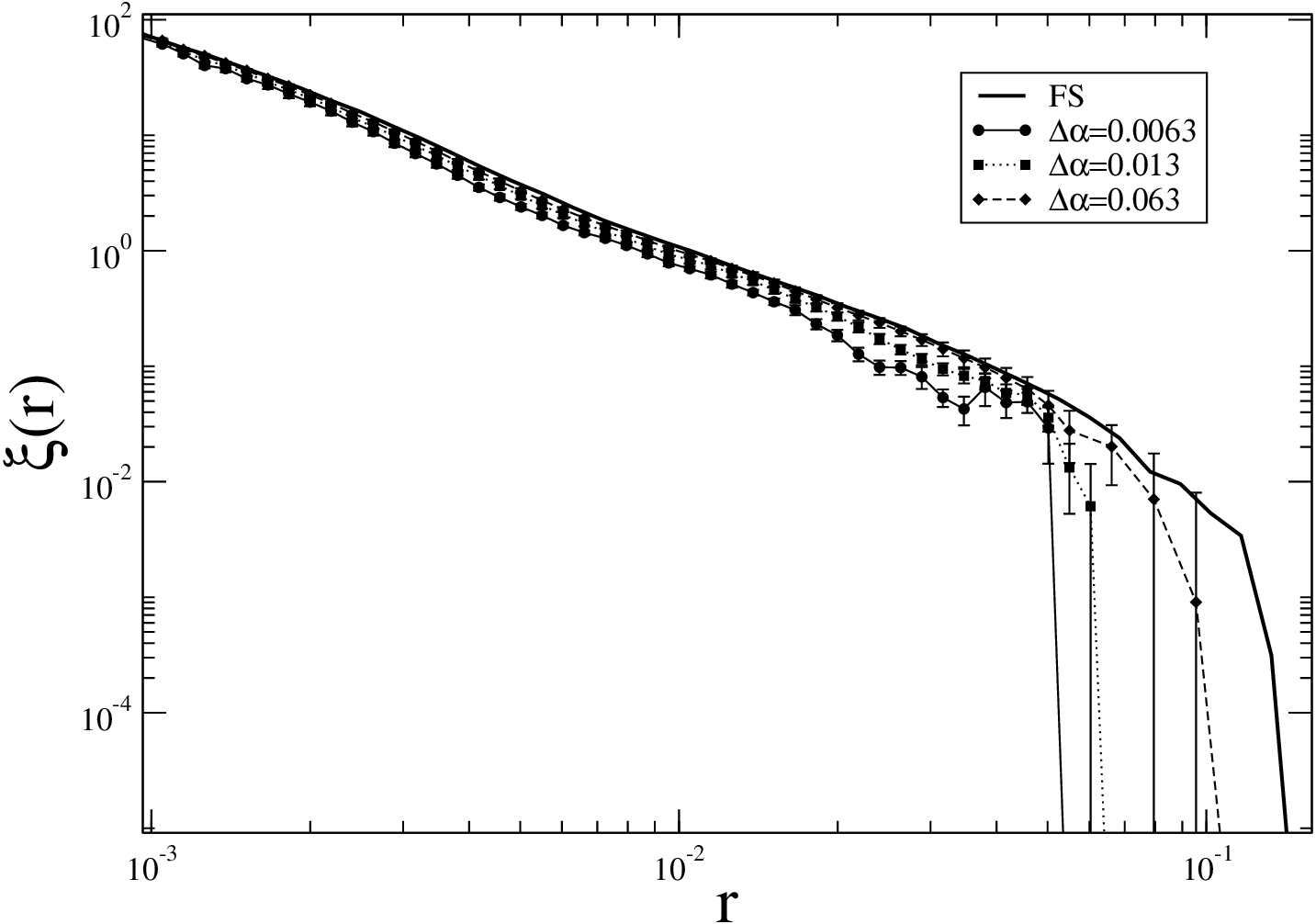} 
\end{center}
\caption{Average 
correlation function computed by using the DP estimator in $N_f=30$
angular slices with $\Delta \alpha = 0.0063, 0.013, 0.063$. The solid
line represents the correlation function computed by using the FS
estimator in independent sub-boxes of side $\ell=0.5$.}
\label{xislice3c}
\end{figure}

In Fig.\ref{xislice4} we finally show the average behavior of the LS
estimator in $N_f=30$ angular slices with $\Delta \alpha = 0.063$ and
with a varying depth of the sample $R_{depth}=0.1,0.2,0.5$. The finite size
dependence of the amplitude and of the scale at which the break in the
power-law behavior occurs is clear. This represents an interesting test 
to be performed in the galaxy data as we discuss below. 

\begin{figure}
\begin{center}
\includegraphics*[angle=0, width=0.5\textwidth]{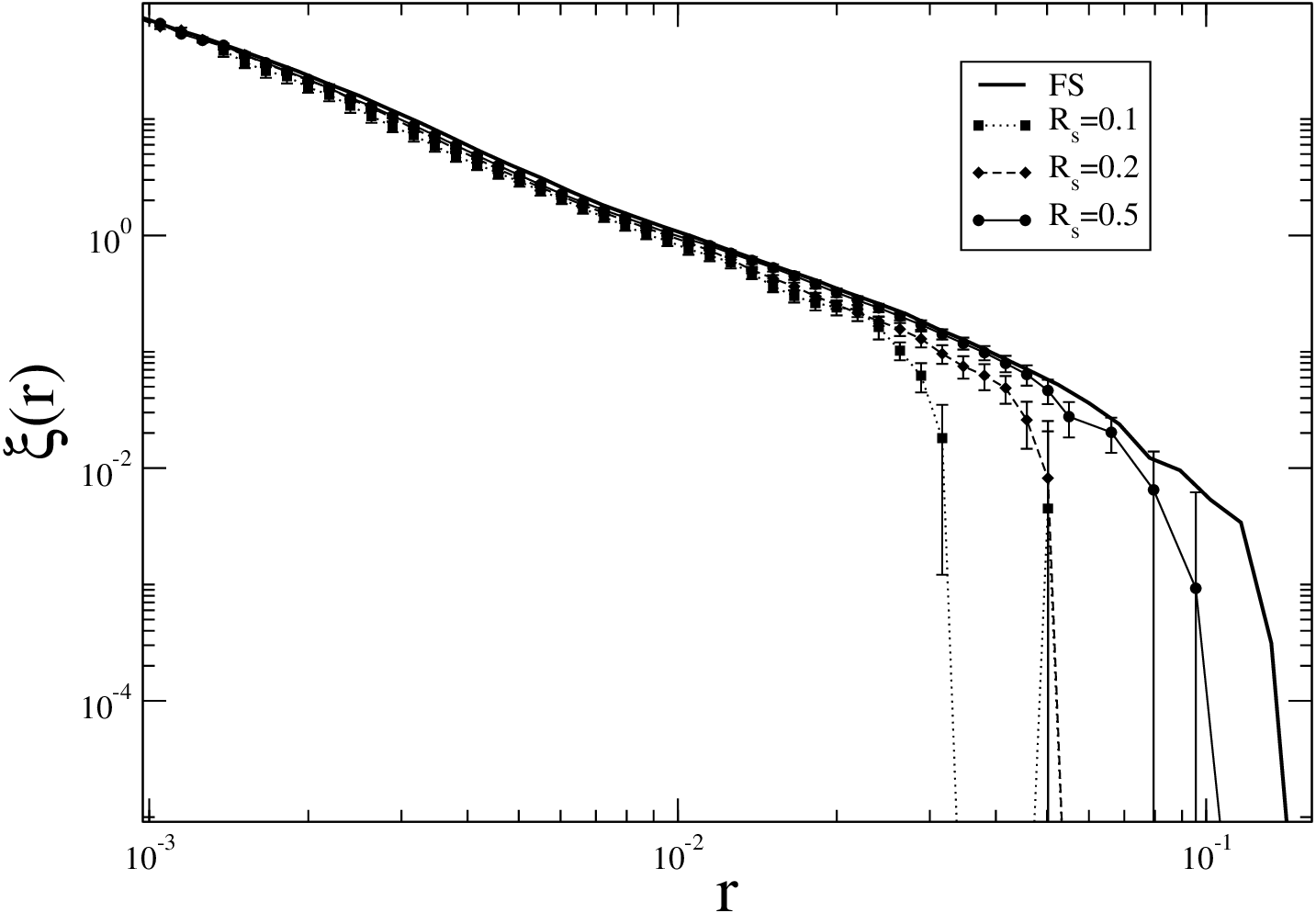} 
\end{center}
\caption{Average 
correlation function computed by using the LS estimator in $N_f=30$
angular slices with $\Delta \alpha = 0.063$ and with a varying depth
of the sample $R_{depth}=0.1,0.2,0.5$. The solid line represents the
correlation function computed by using the FS estimator in independent
sub-boxes of side $\ell=0.5$.}
\label{xislice4}
\end{figure}

\subsection{Determination of the errors}

In Fig.\ref{fig_errors} we show the behavior of the errors computed by
Eq.\ref{err_poiss}, Eq.\ref{err_ftf}, Eq.\ref{jackerrors} and
Eq.\ref{xisigma}. One may note that errors determined by the jackknife
method Eq.\ref{jackerrors} are approximatively the same as the ones computed by the
field-to-field fluctuations Eq.\ref{xisigma}, except at small scales
where the jackknife method is more efficient giving smaller
fluctuations (see discussion in Scranton et al., 2002 and Zehavi et
al., 2002). On the other hand the jackknife error is greater than  the
two other estimators Eq.\ref{err_poiss} and Eq.\ref{err_ftf} (see also
Ross et al., 2006).  Apart the small difference at small scales
between Eq.\ref{jackerrors} and Eq.\ref{xisigma} the former are larger
at scales comparable with the sample size and give a more
conservative estimation of the fluctuations.
\begin{figure}
\begin{center}
\includegraphics*[angle=0, width=0.5\textwidth]{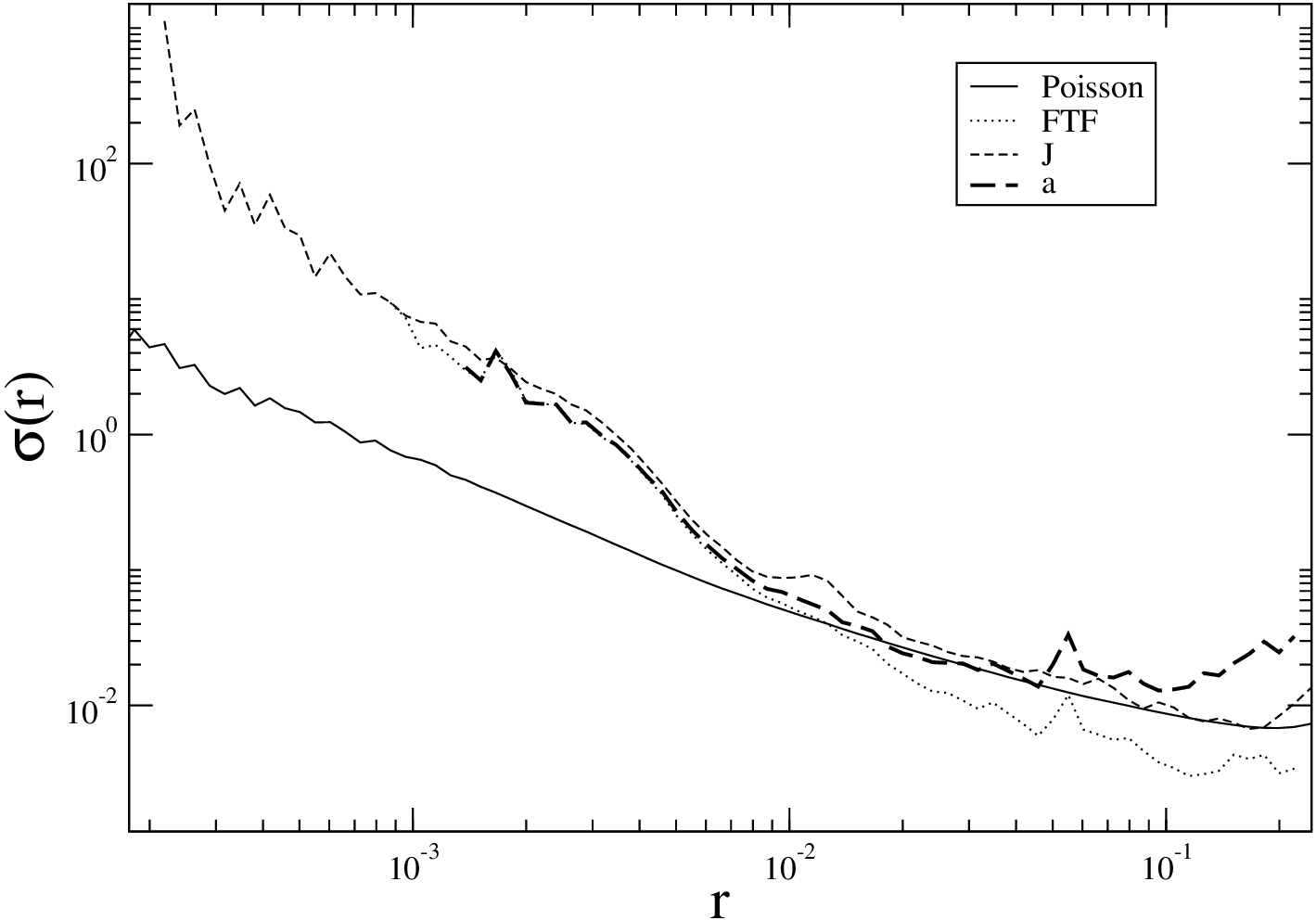} 
\end{center}
\caption{Errors in the estimation of the correlation function
determined by Eq.\ref{err_poiss} (Poisson), Eq.\ref{err_ftf} (FTF),
Eq.\ref{jackerrors} (J) and Eq.\ref{xisigma} (a).}
\label{fig_errors}
\end{figure}

\subsection{Summary and discussion}

We have studied the finite size dependence of the estimated two-point
correlation function by considering three different estimators the FS,
the DP and the LS. We considered the case of a point distribution
presenting, on large enough scale in the sample, weak ($\xi(r)<1$)
power-law correlations. We have performed a series of tests to
establish the role of the biases due to the integral constraint. This
is the principal systematic effect which affects the behavior of the
estimated correlation function at large scales, independently on the
particular estimator considered. Let us briefly discuss our main
results.

We have first considered the determination of the correlation function
in the cubic subsample of size $\ell < L=1$, where $L$ is the whole box
size.  We have constructed our estimation as an average over $N_f$
disjointed sub-boxes. We have studied the behavior of the FS estimator
as a function of the size $\ell$ of the sub-boxes, finding a clear
finite size dependence of both the amplitude (for small $\ell$) and of
the length scale $r^*$ characterizing the break of the power-law
behavior, beyond which the correlation function becomes negative. In
agreement with a simple analytical study of the problem discussed
in the previous section we found that $r^* \sim \ell/2$.  A similar
situation occurs for the LS and DP estimators even though in this case
$r^* \sim \ell$. We note that the LS and DP estimators give very
similar results over the whole range of scales.

In order to understand in more detail the spatial extension of the
reliable measurements of two-point correlations provided by different
estimators we have considered samples with a geometry more similar to
the case of real galaxy samples. Namely we have considered a sphere
around the central point in the box of size $L$ and divided it in
$N_f$ sub-samples with same solid angle $\Omega$. We also considered
subsequent cuts in the depth $\ell<L$. We found that the length-scale
$r^*$ shows a dependence on $\Omega$ and it typically reaches a value
of order of a fraction of $\ell$ which is larger than the scale $R_s$,
up to which the FS estimator can be applied and which is of the
order of the radius of the maximum sphere fully enclosed in the sample
volume.  We have then measured that the scale $r^*$ has a strong
dependence on the value of $\ell$ as for the case of the simple cubic
volumes considered in the previous test.

It is important to note that the tests discussed here have been
performed on a distribution which becomes uniform well inside the
sample size. The above considerations on the performance of the
various estimators can be easily verified for other distributions
which satisfy the property of becoming uniform well inside a given
sample and which show different correlation properties on large
scales.  However the situation is rather different for the case in
which a distribution exhibits strong clustering inside a given sample
without a clear crossover toward a uniform distribution. In this case
the best estimator is the most conservative one, i.e. the FS estimator
as the estimation of the sample density is certainly biased at any
scale as long as the distribution is characterized by strong
non-linear clustering (see discussion in, e.g. Gabrielli et al., 2004
for the treatment of the strongly correlated case).

This situation puts a serious warning on the determination of the
correlation at large scales in a given sample. If an estimator
correlation function presents a break of, for example, the power-law
behavior at a certain scale, the crucial test to be performed is to
check whether this is a finite size or whether it is a true
break. This situation is especially relevant for CDM-type correlations,
for which the correlation function, according to theoretical models,
should present a break from the small-scale power-law correlation at a
scale of order 124 Mpc/h. We will come back on this point in the
conclusion.

Finally we have considered different determinations of the errors of
the estimators of the two-point correlation function. The more
conservative way to estimate errors consists in the computation of the
correlation function in disjointed regions and then to compute the
average and the variance on the average: this method is less efficient
than the jackknife method at small scales but gives similar results to
that at large scales.

\section{Conclusions}

We have considered the real-space properties of CDM density fields,
focusing in detail in a particular variant known as $\Lambda$CDM
(vanilla) model. It is well known that the power-spectrum has
typically a behavior $P(k) \sim k^{m}$ with $-1<m\le-3$ for large
wavelengths $k > k_c$, and $P(k)\sim k$ at smaller wavelengths $k
<k_c$. We discussed that, correspondingly, the two-point correlation
function shows approximatively a positive power-law behavior $\xi(r)
\sim r^{-2}$ at small scales $r < r_c \approx k_c^{-1}$ and a negative
power-law behavior $\xi(r) \propto - r^{-4}$ at large scales $r>r_c$,
where the zero-crossing occurs at about $r_c \approx 124$ Mpc/h in the
model considered. We discussed the fact that, globally, a system with
this type of correlations belong to the category of super-homogeneous
distributions, which are configurations of points more ordered than a
purely uncorrelated (Poisson) distribution.  Correspondingly
fluctuations are depressed with respect to the Poisson case, and the
normalized mass variance, for instance, decay faster ($\sigma^2(r)
\sim r^{-4}$) than for the Poisson case ($\sigma^2(r) \sim
r^{-3}$). The condition of super-homogeneity is expressed by the
condition that $P(k) \rightarrow 0$ for $k \rightarrow 0$, or
alternatively that
\[
\int_0^{\infty}  \xi(r) r^2 dr = 0 \;.
\] 

Following the work of Durrer et al. (2003) we have pointed out that
the above condition is broken when one samples the distribution, as
for example when the simplest biasing scheme of correlated Gaussian
fields (introduced by Kaiser, 1984) is applied. This is particularly
important for the behavior of the power-spectrum for $k<k_c$, which,
under biasing, remains constant instead of going as $P(k) \sim k$. The
correlation function at large scales $r>r_c$ is instead expected to be
linearly amplified with respect to the original one of the whole
matter field. Thus the large scale negative tail $\xi(r) \sim -r
^{-4}$ is the main feature which one would like to detect in order to
test theoretical models.

Given the fact that when $\xi(r)$ becomes negative, it is
characterized by a very small amplitude, the determination of the
negative power-law tail represents a very challenging problem. We have
discussed the fact that, at first approximation in a real measurement,
one may treat the system as having positive correlations at small
scales with an exponential cut-off at the scale $r_c$ and then it
becomes uncorrelated (a situation which can be regarded as upper limit
to the presence of anti-correlations). This implies that for $r_c >
124$ Mpc/h galaxy distribution should not present any positive
correlation. Whether this behavior is compatible with the existences
of structures of order 200 Mpc/h or more is an open problem which has
to be addressed in the studies of forthcoming galaxy catalogs.

More in detail, one of the most basic results (see e.g., Peebles
1980) about self-gravitating systems, treated using perturbative
approaches to the problem (i.e. the fluid limit), is that the
amplitude of small fluctuations grows monotonically in time, in a way
which is independent of the scale.  This linearized treatment breaks
down at any given scale when the relative fluctuation at the same
scale becomes of order unity, signaling the onset of the
``non-linear'' phase of gravitational collapse of the mass in regions
of the corresponding size.  If the initial velocity dispersion of
particles is small, non-linear structures start to develop at small
scales first and then the evolution becomes ``hierarchical'', i.e.,
structures build up at successively larger scales. Given the finite
time from the initial conditions to the present day, the development
of non-linear structures is limited in space, i.e., they can not be
more extended than the scale at which the linear approach predicts
that the density contrast becomes of order unity at the present
time. This scale is fixed by the initial amplitude of fluctuations,
constrained by the cosmic microwave background anisotropies (Spergel et
al., 2007), by the hypothesized nature of the dominating dark matter
component and its correlation properties.  According to current models
of CDM-type the scales at which non-linear clustering occurs at the
present time (of order 10 Mpc) are much smaller than the scale $r_c
\approx 124$ Mpc/h (see e.g. Springel et al., 2005). 
Thus the region where the super-homogeneous
features should still be in the linear regime, allowing a direct test
of the initial conditions predicted by early universe models. The
scale $r_c$ marks the maximum extension of positively correlated
structures: beyond $r_c$ the distribution must be anti-correlated
since the beginning, as there was no time to develop other
correlations.  The possible presence of structures, which mark
long-range correlations, whether or not of large amplitude, reported
both by observations of galaxy distributions (like the Sloan Great
Wall --- see Gott et al., 2005), by the detection of dark matter
distributions (see e.g. Massey et al., 2007) and by the large void of 
radius $\sim 140$ Mpc identified by Rudnick et al. (2007), is
maybe indicating that positive correlations extend well beyond $r_c$.

We have discussed that an important finite size effect must be
considered when estimating the correlation function, and which may
mimic a break of the power-law behavior similar to the ones of CDM
models at a scale of order $r_c$. This is related to the effect 
of the integral constraint in the estimators, namely the fact that 
the sample average, estimated in a finite sample, differs
from the ensemble average, and can be finite-size dependent.
This situation occurs when correlations (weak or strong) extend  
to scales larger than the sample size.

For these reasons, in order to study the two-point correlation
function in real galaxy samples when its amplitude becomes smaller
than unity, it is crucial to check whether the break of the power-law
behavior has a finite size dependence or not, by choosing samples with
different depth. In this perspective the assessment of the reality of
the break of the two-point correlation function is the main
observational point to be considered.  Once this will be clarified
other features should be considered, as for the example the so-called
baryonic bump, which is a very small perturbation to the overall shape
of the correlation function at scales of order of the zero-point
$r_c$. We will present a detailed analysis of the correlation
properties of galaxy distribution in the SDSS catalog, considering
specific tests for finite-size effects in the determination of the
correlation function, in a forthcoming paper.

\section*{Acknowledgments}
We are grateful to Y. Baryshev, A. Gabrielli, M. Joyce, B. Marcos and
L. Pietronero for useful discussions and comments. FSL thanks the
MIUR-PRIN05 project on ``Dynamics and thermodynamics of systems with
long range interactions" for financial support.

{}

\end{document}